\documentclass[aps,prd,10pt,longbibliography,showkeys,floatfix,showpacs,nofootinbib]{revtex4-1}
\usepackage[utf8x]{inputenc}
\usepackage{amsfonts,amsmath, bm,bbold,latexsym,amssymb, graphicx}
\usepackage{hyperref, color}
\providecommand{\U}[1]{\protect\rule{.1in}{.1in}}

\newcommand{\bmath}{\begin{displaymath}}
\newcommand{\emath}{\end{displaymath}}
\newcommand{\bite}{\begin{itemize}}
\newcommand{\eite}{\end{itemize}}

\newcommand{\tr}{{\rm Tr}}
\renewcommand{\o}{\omega}

\newcommand{\eps}{\varepsilon}

\newcommand{\one}{\ensuremath\mathbb{1}}
\newcommand{\half}{{\textstyle \frac{1}{2}}}

\renewcommand{\o}{\omega}
\newcommand{\oP}{{\omega_p}}
\newcommand{\bx}{\mathbf{x}}
\newcommand{\by}{\mathbf{y}}

\newcommand{\bk}{\mathbf{k}}
\newcommand{\bq}{\mathbf{q}}
\newcommand{\br}{\mathbf{r}}

\newcommand{\bP}{\vec{P}}

\newcommand{\bG}{\mathbf{G}}
\newcommand{\bV}{\mathbf{V}}
\newcommand{\bT}{\mathbf{T}}
\newcommand{\bt}{\mathbf{t}}

\newcommand{\beps}{\bm{\varepsilon}}
\renewcommand{\O}{\mathcal{O}}

\newcommand{\doubleint}{\int\hspace{-.9em}\int}

\newcommand{\bel}[1]{\begin{equation}\label{#1}}
\newcommand{\bal}[1]{\begin{eqnarray}\label{#1}}
\newcommand{\ee}{\end{equation}}
\newcommand{\ea}{\end{eqnarray}}
\newcommand{\equ}[1]{~Eq.(\ref{#1})}
\newcommand{\vev}[1]{\langle #1\rangle}

\newcommand{\fig}[1]{~Fig.~\ref{#1}}

\newcommand{\drop}[1]{}

\begin{document}
\title[NCasimir]{Perturbative Roughness Corrections to Electromagnetic Casimir Energies}
\author{Hua Yao Wu and Martin Schaden}
\affiliation{Department of Physics, Rutgers University, 101 Warren Street, Newark NJ 07102}

\begin{abstract}
Perturbative corrections to the Casimir free energy due to macroscopic roughness of dielectric interfaces are obtained in the framework of an effective low-energy field theory. It describes the interaction of electromagnetic fields with materials whose plasma frequency $\oP$ determines the low-energy scale. The na{\"i}ve perturbative expansion of the single-interface scattering matrix in the variance of the profile is sensitive to short wavelength components of the roughness correlation function.  We introduce generalized counter terms that subtract and correct these high-momentum contributions to the loop expansion. To leading order the counter terms are determined by the phenomenological plasmon model. The latter is found to be consistent with the low-energy description. The proximity force approximation is recovered in the limit of long correlation length and gives the upper limit for the roughness correction to the Casimir force. The renormalized low-energy theory is insensitive to the high-momentum behavior of the roughness correlation function. Predictions of the improved theory are compared with those of the unrenormalized model and with experiment. The Casimir interaction of interfaces with low levels of roughness is found to be well reproduced by that of flat parallel plates with the \emph{measured} reflection coefficients  at a distance that is slightly less than the mean separation of the rough surfaces.
\end{abstract}





\startpage{1}
\endpage{120}
\maketitle

\section{Introduction}
Casimir originally\cite{Casimir19481} obtained the force due to electromagnetic zero-point fluctuations between two large ideal parallel metallic flat surfaces at vanishing temperature. His approach was soon generalized to dielectric surfaces\cite{Lifshitz19561,*Lifshitz19562,Schwingerin1978}, finite temperature\cite{Schwingerin1978,Brown19691}, and experimentally more accessible geometries\cite{Derjaguin19581}.  Because it was unimportant in early Casimir experiments\cite{Lamoreauxin1997}, the the influence of surface roughness was investigated only later\cite{Maradudin19801,*Mazur19811, Novikov19901,*Novikov19902,*Novikov19921,*Novikov19922}. Once Casimir forces were accurately measured with atomic force microscope techniques\cite{Mohideen19981} at plate separations of a few hundred nanometers, this correction could no longer be ignored. Effects due to surface roughness are even more important at the small separations and higher accuracy of recent experiments\cite{Decca20031,*Decca20071,Zwol20071,*Zwol20081}. Increasing experimental\cite{Pala2008,Broer2012} and theoretical\cite{Klimchitskaya19991,Emig20011,Emig2004,Neto2005,Genet20031,* Neto20051, *Neto20061, *Lambrecht2006, Palasantzas20051, Broer2012, Bordag2009bk, Mazzi2011, HuaMartinin2012, Mazzitelli2012} effort has since been devoted to understanding this correction. The only rigorous non-perturbative approach to roughness currently is the Proximity Force Approximation (PFA) (and some recent modifications thereof\cite{Broer2012,Bordag2009bk}). This approximation is accurate when the correlation length $l_c$ of the profile greatly exceeds the average plate separation $a$ as well as the inverse plasma frequency $1/\oP$ of the material\cite{Klimchitskaya19991,Bordag2009bk}.  Most rigorous approaches consider perturbative corrections to the Green\rq{}s function in powers of $\sigma/a$ or a derivative expansion of the roughness profile\cite{Mazzi2011}. The limit of very  rough plates with $a\gg l_c$ was first considered in Refs.~\cite{Maradudin19801,*Mazur19811} using methods of stochastic calculus. 

For stochastic roughness all perturbative calculations to date \cite{Maradudin19801,*Mazur19811, Novikov19901,Neto2005,Genet20031,*Neto20051,*Neto20061,Broer2012,Mazzitelli2012} show an increase in magnitude of the Casimir energy and force with decreasing correlation length $l_c$. They  approach the PFA for $l_c \gg a$\cite{Bordag2009bk} as a lower bound.   This behavior corresponds to the dashed curves for the ratio of the roughness correction to the Casimir energy of flat plates shown in\fig{Roughcorr}. As we have argued in\cite{HuaMartinin2012} for a scalar field, such a strengthening of the Casimir force due to roughness is not just counter-intuitive but also unphysical. From the point of view of the multiple scattering expansion, decreasing the correlation length decreases the magnitude of the Casimir force since the reflection coefficient for back-scattering is reduced. Corrections to the free energy of higher order in the loop expansion for scalr fields are of similar magnitude\cite{HuaMartinin2012,Mazzitelli2012} at small $l_c$. Assuming the scalar model to be valid at any scale, the leading contributions in $\sigma^2/(l_ca)$  were resummed\cite{HuaMartinin2012}. In the Casimir energy this  effectively amounted to reducing the separation between two flat plates. The resummed Casimir energy in this approximation indeed decreases with increasing roughness\cite{HuaMartinin2012}.  We will find a similar behavior for the roughness correction to the electromagnetic Casimir force with the present approach.     

The perturbative analysis for electromagnetic fields  in \cite{Maradudin19801,*Mazur19811,Novikov19901,*Novikov19902,*Novikov19921,*Novikov19922,
Genet20031,Neto20051,*Neto20061} also predicts a strengthening of the Casimir force with increasing roughness. This trend does not appear to be supported by experiment\cite{Decca20031,*Decca20071,Pala2008}. Most experimental investigations\cite{Chan2008, Decca2013}using machined uni-directional surfaces are in a non-perturbative regime. However, a perturbative analysis of roughness can be justified for some investigations\cite{Zwol20071,*Zwol20081,Pala2008} that use relatively thin rough gold coatings. The roughness correction compared to flat plates observed at small separations in this case is of the order of $30\%$ only. These experiments  appear to measure a Casimir force that is smaller (not larger) than the PFA estimate.

We here set out to extend the low-energy formalism we developed for scalar fields to  interactions of  the electromagnetic field with matter. Apart from being more complicated, the basic field-theoretic approach is similar.  However, there is a fundamental difference between the effective low-energy theories for scalar and electromagnetic fields: whereas one can pretend that the scalar theory is valid at all length scales, this is not possible for the electromagnetic model.  The coupling to the roughness profile  sets the energy scale in the scalar model, whereas this interaction for the electromagnetic case is dimensionless. The scale of the low-energy effective electromagnetic theory is the plasma frequency, $\oP$. The description of electromagnetic interactions with materials by their dielectric permittivity is not reasonable for momentum and energy transfers much above $\oP$. Explicitly resumming high momentum contributions to the loop expansion thus would be quite out of control in the low-energy effective electromagnetic model.
         
Unfortunately, the roughness contribution to the low-energy electromagnetic scattering matrix arises to a large part from (loop) momenta $q\sim 1/l_c\gtrsim \oP$. Using the low-energy theory to compute these high-momentum contributions is not justified.  Instead of resumming high orders of the loop expansion, we  use generalized counter terms to correct for the high-momentum contributions\cite{Gomis19961,Weinbergbook,Leutwyler1994}. One thus trades (wrong) high-momentum contributions to the scattering matrix for a phenomenological description -- in this case plasmon scattering. 

The observation that UV-divergent vacuum energies arise due to unphysical boundary conditions is quite old\cite{Deutsch:1978sc}. Some UV-divergences may be absorbed in the renormalization of physical parameters\cite{Jaffe2002, *Elizalde2003, *Jaffe2004, *Kirsten2009, *Fulling2011, *Emigknife2011}. Sometimes they are avoided by a more realistic modeling of the surface. However, this invariably gives results that are \emph{sensitive} to the modeling of the interactions with materials at high energies. Here we address the related, but somewhat different issue, that the low-energy description is not suited for computing  high-momentum contributions to physical observables, whether they diverge or not. High momentum parts of loop integrals should not be viewed as reliable predictions of an effective low-energy theory and generally have to be corrected phenomenologically. 

This article is organized as follows. In Sec.~\ref{Roughness} we present Schwinger's low-energy theory for electromagnetic interactions with materials and derive the scattering matrix $\bT^h$ for roughness corrections to the Casimir free energy. The one-loop correction to $\bT^h$ is found to be UV-sensitive in Sec.~\ref{secTh}.  We show that this problem may be solvedby subtraction and inclusion of a phenomenological plasmon contribution. In Sec.~\ref{secIV} roughness corrections to the Casimir free energy to leading order of the variance are derived. They differ from earlier results by the inclusion of a counter term that corrects uncontrolled high-momentum contributions to loop integrals. We obtain the limits of very large and very small correlation length as well as the ideal metal limit and determine the plasmon coupling  at low energies by analyticity arguments.  Sec.~\ref{secV} develops the low-energy effective field theory of electromagnetic interactions with materials to one loop including generalized counter terms. We state the renomalization conditions that determine them. Sec.~\ref{secNum} presents our numerical results and compares them to unrenormalized perturbation theory and experiment. Sec.~\ref{Concl} is a summary of the approach. Basic ingredients and some detailed calculations are relegated to four appendices.

\section{The Electromagnetic Free Energy of a Rough and a Flat Material Interface}
\label{Roughness}
 The present approach is based on Schwinger's low-energy effective field theory\cite{Schwingerin1978} for electromagnetism. The partition function in this model is a functional of the local dielectric permittivity tensor $\beps(\zeta_n,\bx,z)$ of the material and of an external polarization source $\vec{P}_n(\vec{x})=\vec{P}(\zeta_n, \vec{x})$.  It is the product of contributions from (independent) thermal modes\cite{Fried1972bk,*Becher1984bk,*Kapusta1989bk} of the electric field to Matsubara frequency\footnote{We adopt natural units $\hbar=c=k_B=1$ and  suppress the index $n$ of $\zeta_n$ in summations $\sum_n$ over all Matsubara frequencies. } ,
\bel{Matsubara}
\zeta_n=2\pi |n| T\ge0  \ \text{for all} \ n\in\text{Integers}\ .
\ee
The partition function formally is given by  the functional integral,
\bel{SchwingerFn}
Z_T[\vec{P};\beps]\propto\prod_n\int D[\vec{E}_n]\exp\left\{-\frac{1}{2T}\int d^3x \vec{E}_n^\dagger(\vec{x})[\beps(\zeta_n,\vec{x})+\frac{1}{\zeta_n^2}\nabla\times\nabla\times] \vec{E}_n(\vec x)+2 T \vec{E}_n(\vec{x})\cdot \vec{P}_n(\vec{x})\right\}\ .
\ee
$Z_T[\vec{P};\beps]$ is the partition function of QED in axial gauge $A_0=\Phi=0$ for a medium with local dielectric permittivity $\beps(\zeta,\vec x)$. In this gauge $\vec{E}_n=\zeta\vec{A}_n, \vec{B}_n=\nabla\times \vec{A}_n$ and the  current source  $\vec{j}_n=\zeta \vec{P}_n$. 

\begin{figure}
\includegraphics[scale=0.5]{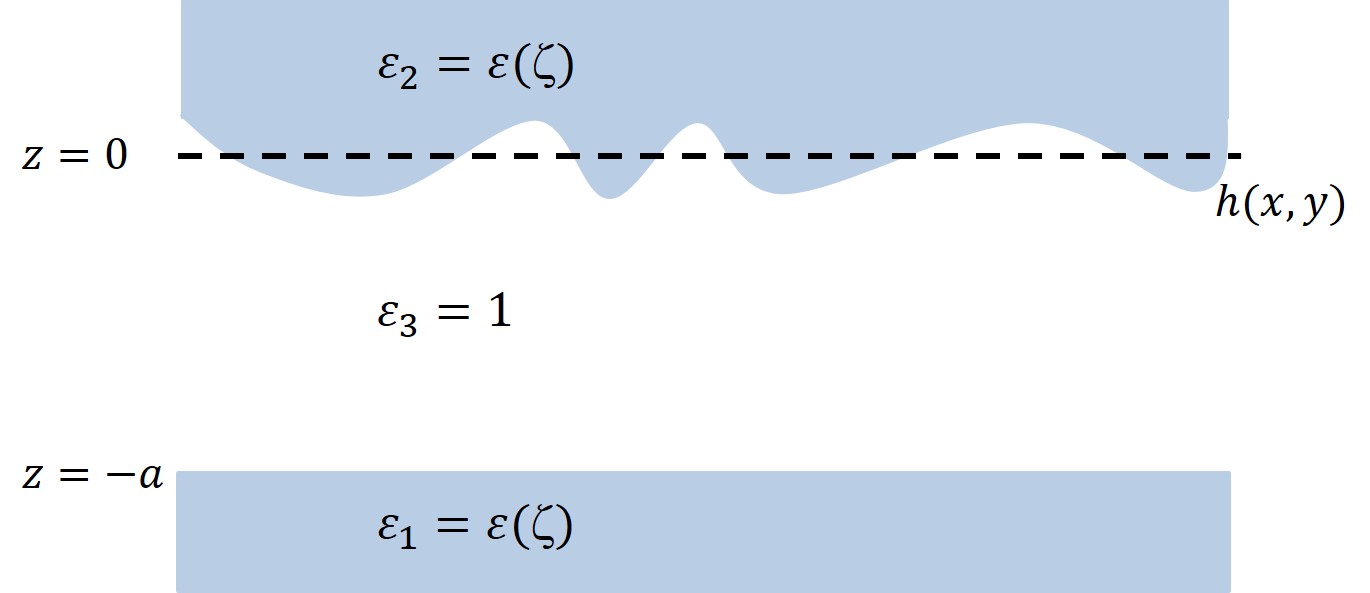}
\caption{Two semi-infinite slabs of the same material separated by vacuum. The low-energy electromagnetic properties of the material are described by a bulk-permittivity $\eps(\zeta=i\omega)$ that only depends on the frequency of the electric field.  In Cartesian coordinates the planar interface is at $z=-a$ and the mean separation of the two interfaces is $a$. The surface of the rough slab is at $z=h(\bx)$ where $h(\bx)$ is a profile function that generally depends on both transverse coordinates $\bx=(x,y)$. We develop a perturbative expansion valid for $|h(\bx)|\ll a$ with no restrictions on the profile other than that it be single-valued. $h(\bx)$ in particular need not be as smooth as shown here.} 
\label{slabs}
\end{figure}

We consider the standard Casimir configuration of two parallel semi-infinite plates at an average separation $a$ that is much less than their transverse dimension\cite{Casimir19481}.  In the following we restrict the discussion to the configuration shown in\fig{slabs} of two semi-infinite dielectric (metallic) slabs of the same material  separated by vacuum, only one of which is rough,  
\bel{epsdef} 
\eps_3(\zeta)=1\ ,\ \ \eps_2(\zeta)=\eps(\zeta)=\eps_1(\zeta)\ .
\ee
We forego the ability to address lateral Casimir forces. Which are finite and vanish if one of interfaces is flat.  Lateral Casimir forces depend on cross-correlations of the two profiles. At separations $a\gg\sigma$ they are small and involve only low momenta.  For corrugated plates they have been computed in \cite{Emig20032, *Cavero20081, *Cavero20082}. We here are interested in the effect of profiles on the \emph{normal} Casimir force. The physical interpretation and consistent subtraction of (potentially divergent) contributions will be our main concern.

The rough interface is assumed to be without enclosures and the deviation from a flat one at $z=0$ is described by a single-valued function $h(\bx)$  that satisfies\footnote{A Cartesian coordinate system with $z$-axis normal to the plates is used to describe this system. We use bold type $\bx=(x,y)$ for 2-dimensional vectors perpendicular to the $z$-axis, whereas $\vec v$ denotes an ordinary 3-dimensional vector.},
\bel{vev0}
\vev{h}:=A^{-1}\int_A d\br\; h(\br)=0 . 
\ee
The point of reference for defining the separation $a$ of the two slabs should be irrelevant. However, a  consistent perturbative expansion is feasible only in the absence of so-called tadpole contributions. These vanish if the separation $a$ is such that \equ{vev0} holds. \equ{vev0} in this sense defines the distance $a$ between the interfaces.

When the cross-sectional area $A$ of the slab is taken arbitrary large, boundary effects can be ignored and the 2-point correlation function,
\bel{defD2}
D_2(\bx-\by)=\vev{h(\bx) h(\by)}:=A^{-1}\int_A d\br\; h(\br+\bx) h(\br+\by)
\ee
is invariant under transverse translations. The roughness variance, 
\bel{defvar}
\sigma^2=D_2(0),
\ee
is a measure for the roughness amplitude.   
  
The dielectric permittivity function $\beps(\eps,\vec{x})$ in this effective low-energy field theory is of the form, 
\begin{align}\label{eps}
\beps(\zeta,\vec{x})&=\one[\eps_3(\zeta)+(\eps_2(\zeta)-\eps_3(\zeta))\theta(z-h(\bx))+(\eps_1(\zeta)-\eps_3(\zeta))\theta(-z-a)]\nonumber\\&=\bV^\parallel(\zeta,z)+\bV^h(\zeta,\bx,z)\ ,
\end{align}
where
\bel{defVh}
\bV^h(\zeta,\bx,z)=\one[(\eps_2(\zeta)-\eps_3(\zeta))(\theta(z-h(\bx))-\theta(z))]
\ee
is the deviation due to the roughness profile $h(\bx)$ from the dielectric permittivity of a transversely homogeneous medium given by,
\bel{Vslabs}
\bV^\parallel(\zeta,z)=\one[\eps_3(\zeta)+(\eps_2(\zeta)-\eps_3(\zeta))\theta(z)+(\eps_1(\zeta)-\eps_3(\zeta))\theta(-z-a)]+\delta \bV^{h}(\zeta,z)\ .
\ee
We shall argue that the counter term $\delta\bV^h(\zeta,z)$ to the dielectric permittivity of three flat slabs is necessary for a consistent perturbative expansion in the framework of a low-energy theory.  $\delta\bV^h(\zeta,z)$ depends on gross properties of the profile $h(\bx)$ but not on the transverse position $\bx$ nor on the separation of the two interfaces.  This counter term ensures that the single-interface scattering matrix is reproduced by the low-energy theory.   To leading order $\delta\bV^h(\zeta,z)$ is proportional to the variance $\sigma^2$ of the rough interface. We are thus calculating the perturbative expansion for the rough interface at $z=0$ about an effective $\bx$-independent (bare) permittivity, 
\bel{epseff}
\bm{\eps}_\text{eff}(\zeta,z)=\one\eps_2(\zeta)\theta(z)+\one\eps_3(\zeta)\theta(-z)+\delta\bV^{h} (\zeta,z)\ .
\ee
 $\delta\bV^h (\zeta,z)$ has support near the surface at $z\sim 0$ only\footnote{To first order in the variance we find in \equ{dtV} of Sect.~\ref{secTh} that $\delta\bV^h(\zeta,z)\propto \delta(z)$, that is  $\bm{\eps}_\text{eff}(\zeta,z)$ differs from that of an interface by the insertion of an arbitrary thin plate.}. To approximate scattering off a rough interface by an effective $\bm{\eps}_\text{eff}(\zeta,z)$  is a conceptually appealing idea and not new \cite{RosaDalvit2008,Gusso2012}. We develop a consistent low-energy approach in which this is realized perturbatively. Contrary to commonly used ans{\"a}tze for the effective $\bm{\eps}_\text{eff}$,   $\delta\bV^h(\zeta,z)$  generally is not isotropic. 

The inherent limitations of the effective low-energy description derive from the fact that electromagnetic interaction with matter is encoded in the permittivity function. They are not restricted to a perturbative analysis. The dimensionless permittivity $\eps(\zeta)=\eps(\zeta/\oP)$ depends implicitly on a scale that can be identified with the plasma frequency $\oP$ of the material.  At momentum- or  energy-transfers (or temperatures) that are much larger than $\oP$ the effective  low-energy theory of\equ{SchwingerFn} fails to incorporate non-linear effects or to account for the creation of free charges. 
The ansatz that the permittivity does not depend on the profile furthermore is incorrect at wavelengths comparable to the plasma wavelength $l_p=2\pi/\oP$ and a description in terms of the bulk permittivity of the homogeneous material is not warranted within the plasma skin depth of order $l_p$. For gold surfaces commonly used $\oP\sim 0.046\text{nm}^{-1}\sim 9\text{eV} $. The low-energy description of electromagnetic interactions with such materials by \equ{eps} therefore is already  questionable at wave numbers $q\gg \oP\sim 0.046\text{nm}^{-1}$ that  resolve less than $20\text{nm}$ or about $200$ gold atoms. We will find that roughness corrections to the Casimir force with correlations lengths  $l_c\lesssim 1/\oP$  depend on momentum transfers $q\gg \oP$ that are inadequately described by the low-energy theory. The conservative approach is to use the effective low-energy theory to only compute roughness corrections with $l_c\gg 1/\oP\sim 20\text{nm}$, a regime where the PFA generally is quite accurate. We improve on this by  introducing phenomenological input.
  
It is interesting in this regard that many comparisons  of theory with experiments in the literature  are for correlation lengths $l_c\sim 25\text{nm}\sim 1/\oP_\text{Au}$. The unimproved theory is highly sensitive to $l_c$ in this regime and (re)produces large variations with only small changes in parameters.  Roughness corrections computed with this unimproved model for such short correlation lengths are uncontrolled and in fact physically untenable\cite{HuaMartinin2012}.

We herre compute roughness corrections that are consistent with the low-energy effective model by using low-energy (experimental) data to systematically subtract and correct high-momentum contributions to the loop expansion. The method is quite general\cite{Weinbergbook,Gomis19961}  and has been successfully applied to low energy effective field theories  as diverse as chiral perturbation theory\cite{Leutwyler1994}  and (quantum) gravity\cite{Weinbergbook}. In our case it yields a consistent expansion in $\sigma/a$ for any value of $0<l_c \oP<\infty$ at the expense that the reflection of electromagnetic radiation perpendicular to the rough plate has to either be measured or be reliably modeled.   

\subsection{The Green's Function and Casimir Energy of Two Parallel Flat Interfaces}
Schwinger obtained the free energy and the response to an external polarization source $\vec{P}_n(\bx,z)=\vec{P}(\bx,z;\zeta_n)$ for three parallel slabs in the framework of the low energy effective field theory given by \equ{SchwingerFn}. The free energy in this case is\cite{Schwingerin1978},
\begin{align}\label{Fn0}
{\cal F}_T^\parallel(a, \vec{P})=F_T^\parallel(a)-\frac{T^2}{2} \sum_n\{\vec{P}_n|\bG^{\parallel(n)} |\vec{P}_n\}\ .
\end{align}
Here $F_T^\parallel$ is the well-known Casimir free energy for three parallel slabs,
\bel{CasF}
F_T^\parallel(a)=\frac{A T}{2}\sum_n\int \frac{d\bk}{(2\pi)^2} [\ln(1-r_1 r_2 e^{-2\kappa_3 a})+\ln(1-\bar r_1 \bar r_2  e^{-2\kappa_3 a})]\ ,
\ee
where the reflection coefficients at the $i$-th interface of area $A$ for the TE- and TM-modes are,
\bel{rs}
r_i=r_i^{TE}=\frac{\kappa_3-\kappa_i}{\kappa_3+\kappa_i} \ \ \text{and}\ \ \bar r_i=r_i^{TM}=\frac{\bar\kappa_3-\bar\kappa_i}{\bar\kappa_3+\bar\kappa_i}\ ,\ \text{with}\ \  
\kappa_i=\sqrt{\bk^2+\zeta^2\eps_i(\zeta)} \ \ \text{and} \ \ \bar\kappa_i=\frac{\kappa_i}{\eps_i(\zeta)}\ .
\ee

The response to the $n$-th Matsubara mode of an external source of polarization is,
\begin{subequations}
\label{resp}
\begin{align}
\{\vec{P}_n|\bG^{\parallel(n)} |\vec{P}_n\}&=\int dz dz' d\bx d\by  \vec{P}_n^\dagger(\bx,z)\cdot \bG^\parallel(\bx,z,\by, z';\zeta_n, a) \cdot\vec{P}_n(\by,z)\label{respa}\\
&=\int \frac{d\bk}{(2 \pi)^2}dz dz'   \vec{P}_n^\dagger(\bk,z)\cdot \bG^\parallel(\bk,z,z';\zeta_n, a) \cdot\vec{P}_n(\bk,z)\ .\label{respb} 
\end{align}
\end{subequations}
 $\bG^\parallel$  in \equ{respa} is the Green's dyadic  solving\footnote{ $\bG^\parallel(\zeta)$ is related to Schwinger's\cite{Schwingerin1978} dyadic $\mathbf{\Gamma}(\omega)$ at angular frequency $\omega$ by $\bG^\parallel(\zeta)=-\mathbf{\Gamma}(i\omega)$.},
\bel{Gdef}
\Big{[}\bV^\parallel(\zeta,z)+\frac{1}{\zeta^2}\nabla\times\nabla\times\Big{]}\bG^\parallel(\bx,z,\by, z';\zeta,a)=\one\delta(z-z')\delta(\bx-\by)\ .
\ee
Due to translational invariance in transverse directions, $\bG^\parallel(\bx,z,\by,z';\zeta, a)$ is a function of $\bx-\by$ and the Fourier-representations in \equ{respb} are,
\bel{Gk}
\bG^\parallel(\bx,z,\by, z';\zeta,a)=\int \frac{d\bk}{(2 \pi)^2} e^{i\bk (\bx-\by)} \bG^\parallel(\bk,z,z';\zeta, a) \ \text{and}\ \  \vec{P}_n(\bk,z)=\int d\bx\, e^{-i\bk\bx}\vec{P}_n(\bx,z)\ .
\ee
$\bG^\parallel$ can be decomposed into a single-interface Green's dyadic $\bG^|(\bx-\by,z,z';\zeta)=\bG^\parallel(\bx,z,\by,z';\zeta, a\rightarrow\infty)$ where the second interface has been removed and the correction $\bG^{|a|}(\bx-\by,z,z';\zeta, a)$ due to the presence of a second flat interface at mean  separation $a$. In momentum space the latter vanishes exponentially for $a\rightarrow\infty$,
\bel{decompose}
\bG^\parallel(\bk,z,z';\zeta, a)=\bG^|(\bk,z,z';\zeta)+\bG^{|a|}(\bk,z,z';\zeta)\ .
\ee
Explicit expressions for components of $\bG^|(\bk,z,z';\zeta)$ and $\bG^{|a|}(\bk,z,z';\zeta)$ when $z$ and $z'$ are in slab~$\#2$  or slab~$\#3$ are collected in App.~\ref{appA}. 

\subsection{Perturbative Roughness Correction to the Casimir Free Energy: Greens Function Formalism}
A straightforward perturbative expansion in the roughness potential $\bV^h$ is possible only for media with $\eps_2-\eps_3\ll 1$. Since the Casimir free energy itself is rather small, roughness corrections are not very important in this weak coupling scenario. However, the support of $\bV^h$ is restricted to $|z|\leq\text{max}_{\bx}|h(\bx)|\sim\sigma\ll a$ and a perturbative expansion in $\sigma/a$ may exist even for media whose permittivity is rather large. This expansion in fact is possible even for ideal metals. 

The part of the free energy that captures the dependence on the average separation $a$ of two interfaces is by definition the Casimir free energy due to their interaction \footnote{It vanishes in the limit $a\rightarrow \infty$.  At $T=0$ this "free energy" is the Casimir energy and need not vanish.}. In terms of the Greens-dyadic $\bG^\parallel$ of three parallel slabs satisfying \equ{Gdef}, the full Greens-dyadic $\bG$ for the combination of a rough and a flat interface formally is the solution of,
\bel{Gfull}
\Big{[}\one +\bV\bG^\parallel\Big{]}\bG^{\parallel-1}\bG=\one\ ,
\ee
with,
\bel{defbV}
\bV=\bV^h+\delta\bV^h=\one(\eps(\zeta)-1)(\theta(z-h(\bx))-\theta(z)) +\delta\bV^h(\zeta,z)\ .
\ee 
The change in free energy due to roughness of one interface therefore is\cite{Cavero20081,*Cavero20082, Shajesh20111}, 
\bel{DeltaF1}
\Delta F_T[h,a]=-\half\tr \ln(\bG^{\parallel-1}\bG)=\half\tr \ln(\one +\bV\bG^\parallel)\ ,
\ee
where the trace includes a summation over Matsubara frequencies and over a complete set of scattering states. The expression in \equ{DeltaF1} is rather formal because it includes the change in free energy due to roughness in the absence of the second (flat) interface. This infinite single-body contribution to the free energy does not depend on the mean separation $a$. Subtracting from $\Delta F_T[h,a]$ its value when the two interfaces are infinitely far apart gives the correction to the Casimir free energy due to roughness of an interface as,
\begin{align}\label{FC}
\Delta F_T^\text{Cas}[h,a]&:=\Delta F_T[h,a]-\Delta F_T[h,\infty]=\half\tr \ln(\one +\bV\bG^\parallel)-\half\tr \ln(\one +\bV\bG^{|}) \nonumber\\ 
&=\half\tr \ln(\one +\bT^h\bG^{|a|})\ ,
\end{align}
where
\bel{Tdef}
\bT^h=\bV-\bV \bG^| \bT^h
\ee
is the formal scattering matrix due to the roughness potential $\bV$. $\bT^h$ does not depend on the separation $a$ and describes scattering due to roughness in the \emph{absence} of the second (flat) interface.  Since high momenta are exponentially suppressed in $\bG^{|a|}$, the Volterra series of $\Delta F_T^\text{Cas}[h,a]$ in powers of $\bT^h$,
\bel{expandF}
\Delta F_T^\text{Cas}[h,a]=\half\tr \ln(\one +\bT^h\bG^{|a|}) \sim \half\tr\Big{[}  \bT^h\bG^{|a|}-\half\bT^h\bG^{|a|}\bT^h\bG^{|a|}+\dots
\ee
converges when the norm of $\bT^h \bG^{|a|}$ is bounded and sufficiently small. 

\section{The Roughness Scattering Matrix $\bT^h$}
\label{secTh}
Noting that the component $G^|_{zz}(\bk,z,\by,z';\zeta)$  in \equ{gamma} includes a $\delta$-function singularity, \equ{Tdef} can be rewritten,
\bel{Tdef2}
\bT^h=\tilde\bV-\tilde\bV \tilde\bG^| \bT^h\ ,
\ee 
in terms of the Green's dyadic $\tilde\bG^|$ with Fourier components,
\bel{Gtilde}
\tilde\bG^{|}(\bk,z,z';\zeta)=\bG^{|}(\bk,z,z';\zeta)-\text{diag}(0,0,\frac{1}{\eps_z}\delta(z-z'))\ ,
\ee
and a new potential $\tilde\bV$.
$\tilde\bG$ is devoid of $\delta$-function singularities (but not continuous at $z=0$) with components are given in\equ{Ginf}. To order $\sigma^2$ the potential $\tilde\bV$ is,
\bel{tV}
\tilde\bV(\bx,z;\zeta)=\tilde\bV^h(\zeta,\bx,z)+\delta \tilde\bV^h(\zeta,z)=(\eps-1)[\theta(z-h(\bx))-\theta(z)]\,\text{diag}[1,1,\eps\,\theta(z)+\theta(-z)/\eps]+\delta \tilde\bV^h(\zeta,z)\ .
\ee
The reformulation of \equ{Tdef} in the form of \equ{Tdef2} resums local contributions of the same order in $h$. It allows the formulations of a consistent perturbative expansion in $\sigma$ even in the ideal metal limit $\eps(\zeta)\rightarrow\infty$. Just as for $\bV$, the support of $\tilde \bV$ is restricted to to the interval $|z|<\max_\bx |h(\bx)|\sim\sigma$ only. Since $\tilde \bG^{|}$ is free of ultra-local $\delta$-function singularities,  contributions to $\bT^h$ of $n$-th order in $\tilde \bV$ are at least of $n$-th order in the standard deviation $\sigma$ of the profile $h$.    

To second order in $\sigma$ we need only consider the first two  terms of the Volterra series,
\bel{bT2}
\bT^h\approx \tilde\bV-\tilde\bV \tilde\bG^| \tilde\bV\approx\tilde\bV-\tilde\bV^h \tilde\bG^| \tilde\bV^h=\bT^{(1)}+\bT^{(2)}\ ,
\ee
since the counterterm potential $\delta\tilde\bV^h$ is itself of order $\sigma^2$ (as will be seen).
The second-order contribution $\bT^{(2)}$ of \equ{bT2} is at least of order $\sigma^2$ and its integrated expectation to this order is,
\begin{align}\label{vevT2}
\bt^{(2)}(\bx-\by,\zeta)&:=\vev{\int dz\,dz'\,\bT^{(2)}(\bx,z,\by,z';\zeta)}
=-\vev{\int dz\, dz'\, \tilde\bV^h(\bx,z;\zeta)\tilde \bG^|(\bx-\by,z,z';\zeta)\tilde\bV^h(\by,z';\zeta)}+{\cal O}(\sigma^3)\ .
\end{align}
Because $\int dz\tilde\bV^h(\bx, z,\zeta)$ already is of order $\sigma$, the Fourier components of $\bt^{(2)}$ are\footnote{$\bk=(k,0,0)$ here defines the positive $x$-axis and $(k',\theta)$ are polar coordinates of $\bk'$. Note that a (randomly) rough profile preserves translational (and rotational) invariance on average. The \emph{average} scattering matrix of \equ{vevT2} therefore is diagonal in transverse momentum space.},
\begin{align}\label{tk}
t_{xx}^{(2)}(k,\zeta)&=-(\eps-1)^2\int \frac{d\bk'}{(2\pi)^2}\left( \frac{\kappa'\kappa'_\eps \cos^2\!\theta }{\kappa'_\eps+\eps\kappa'}+\frac{\zeta^2\sin^2\!\theta }{\kappa'_\eps+\kappa'}\right) D(q)+{\cal O}(\sigma^3)\nonumber\\
t_{yy}^{(2)}(k,\zeta)&=-(\eps-1)^2\int \frac{d\bk'}{(2\pi)^2} \left( \frac{\kappa'\kappa'_\eps\sin^2\!\theta }{\kappa'_\eps+\eps\kappa'}+\frac{\zeta^2\cos^2\!\theta }{\kappa'_\eps+\kappa'}\right)D(q)+{\cal O}(\sigma^3)\\
t_{zz}^{(2)}(k,\zeta)&=-(\eps-1)^2\int \frac{d\bk'}{(2\pi)^2} \frac{-k'^2}{\eps\kappa'+\kappa'_\eps}(\eps D_{++}(q)+ D_{-+}(q)+ D_{+-}(q)+\frac{1}{\eps} D_{--}(q)) +{\cal O}(\sigma^3)\nonumber\\
t_{xz}^{(2)}(k,\zeta)&=-(\eps-1)^2\int \frac{d\bk'}{(2\pi)^2}\frac{i k'\cos\theta  }{\kappa'_\eps+\eps\kappa'} (\kappa'_\eps D_{++}(q)+\bar\kappa'_\eps D_{+-}(q)-\eps\kappa' D_{-+}(q)-\kappa' D_{--}(q))+{\cal O}(\sigma^3)\nonumber\\
t_{zx}^{(2)}(k,\zeta)&=-(\eps-1)^2\int \frac{d\bk'}{(2\pi)^2}\frac{-i k'\cos\theta  }{\kappa'_\eps+\eps\kappa'} (\kappa'_\eps D_{++}(q)-\eps\kappa' D_{+-}(q)+\bar\kappa'_\eps D_{-+}(q)-\kappa' D_{--}(q))+{\cal O}(\sigma^3)\nonumber\ ,
\end{align}
with $q^2=(\bk-\bk')^2=k^2 +k'^2-2 k k' \cos\theta$. Since the $\tilde G^|_{zx}$ and $\tilde G^|_{zz}$ components of the dyadic (see Appendix~\ref{appA}) are discontinuous at $z=0$, one has to separately consider correlators of positive and negative components of the roughness profile in \equ{tk}. With $ h_\pm(\bx)=h(\bx)\theta(\pm h(\bx))$, these signed correlators are,
\bel{Ds}
D_{\pm +}(q)=D_{\mp -}(q):=\int d\bx e^{i\bq(\bx-\by)}\vev{h_\pm(\bx) h_+(\by)}\ , \ \  D(q)=2 D_{++}(q)+2 D_{+-}(q) =\int d\bx e^{i\bq(\bx-\by)}\vev{h(\bx)h(\by)}\ .
\ee
$D(q)$ is the Fourier transform of the two-point correlation function $D_2(\bx-\by)$ of \equ{def-D}.  As shown in Appendix~\ref{appB}  the signed correlators for a Gaussian generating functional of roughness correlations also are related to the two-point correlator $D_2$ as,
\begin{align}\label{Dsigned}
\vev{h_+(\bx)h_+(\by)}=\vev{h_-(\bx)h_-(\by)}&=\frac{\sigma^2}{2\pi}(\sin\phi+(\pi-\phi)\cos\phi)\\
\vev{h_+(\bx)h_-(\by)}=\vev{h_-(\bx)h_+(\by)}&=\frac{\sigma^2}{2\pi}(\phi\cos\phi-\sin\phi), \ \text{ with }  0\leq\cos\phi=D_2(\bx-\by)/D_2(0)\leq 1\ ,\nonumber
\end{align}
for a roughness correlation function $D_2(r)$ that is positive and monotonically decreasing with $r=|\bx-\by|$.  The signed correlators do not vanish and approach $\pm\sigma^2/(2\pi)$ for $r\rightarrow \infty$ if $D_2(r\sim\infty)\sim 0$. At small separations $r=|\bx-\by|\ll l_c$,  $\cos\phi = D_2(r)/D_2(0)\sim 1-\beta r^\alpha$. Thus $\phi\propto r^{\alpha/2}$ for $r\sim 0$ with an exponent $\alpha> 0$. The expressions of \equ{Dsigned} for small $\phi$ then imply the behavior,
\begin{align}\label{asympcorr}
\vev{h_+(\bx)h_+(\by)}=\vev{h_-(\bx)h_-(\by)}\sim \half D_2(r)\ ; \ 
  \vev{h_+(\bx)h_-(\by)}=\vev{h_-(\bx)h_+(\by)}\sim  -\frac{\sigma^2}{6\pi}(2\beta r^\alpha)^{3/2}  \ \ \text{for}\ r\ll l_c\ .
\end{align}
After Fourier transformation the asymptotic behavior at large momenta $q l_c \gg 1$ of $D_{++}(q)=D_{--}(q)$ is the same as that of $\half D(q\gg 1/l_c)$, whereas the mixed correlations $D_{+-}(q)=D_{-+}(q)$ fall off more rapidly.    

For $l_c\ll 1/\oP$ high momentum contributions are appreciable or even dominate the 1-loop corrections to the diagonal components of the scattering matrix in \equ{tk}. For example,
\bel{txx0}
 t_{xx}^{(2)}(0,\zeta)=-(\eps-1)^2\int_0^\infty \frac{k dk}{4\pi}\left( \frac{\kappa\kappa_\eps }{\kappa_\eps+\eps\kappa}+\frac{\zeta^2 }{\kappa_\eps+\kappa}\right) D(k)\xrightarrow{l_c\sim 0} -\frac{ (\eps-1)^2}{1+\eps}\int_0^\infty \frac{k dk}{4\pi}  k D(k)
\ee
Whether or not loop integrals like \equ{txx0} diverge depends on the roughness correlation function. For Gaussian correlations the integral converges,
\bel{GaussD} 
\vev{h(\bx)h(\by)}=\sigma^2 e^{-\half(\bx-\by)^2/l^2_c}\Rightarrow D_\text{Gauss}(q)=2 \pi\sigma^2 l_c^2 e^{-\half q^2 l_c^2}\ \ \text{with}\ \  \int_0^\infty \frac{k dk}{4\pi}  k D_\text{Gauss}(k)=\frac{\sigma^2}{2 l_c}\sqrt{\frac{\pi}{2}}\ ,
\ee
but the roughness "correction" becomes (arbitrary) large for $l_c \sim 0$. This invalidates the perturbative expansion in $\sigma/a$ and, for sufficiently small $l_c$, violates unitarity. It furthermore is unphysical that roughness corrections to the scattering matrix with profiles of fixed variance become arbitrary large as $l_c\rightarrow 0$.

For a scalar field and Gaussian roughness correlation, higher orders in the loop expansion are of the same order in $\sigma/l_c$  in this limit\cite{HuaMartinin2012}. Assuming the scalar model is valid at all energy scales, we resummed the leading $\sigma/l_c$ contributions to the scalar Casimir energy and found that they amount to a change in the effective separation $\Delta a\sim \sigma^2/l_c$ of the two interfaces.

However, the effective low-energy electromagnetic theory of \equ{SchwingerFn} evidently is not valid for momenta that far exceed the plasma frequency $\oP$. One furthermore is not assured  that summing incorrect higher loop contributions in this effective low energy theory improves the situation. We therefore will not follow that line and proceed differently in this case.

For correlation functions with non-vanishing slope at $r=|\bx-\by|=0$, that is $D'_2(r=0)\neq 0$, the situation is even more serious.  For instance, the 2-dimensional Fourier transform of an exponential correlation function,
\bel{ExpD}
\vev{h(\bx)h(\by)}=\sigma^2 e^{-|(\bx-\by)|/l_c}\Rightarrow D_\text{Exp}(q)=\frac{2\pi\sigma^2 l_c^2}{(1+q^2 l_c^2)^{3/2}}, 
\ee
decays as a power law proportional to $q^{-3}$ at large momenta. The integral in \equ{txx0} and other (diagonal) components of the roughness correction $\bt^{(2)}$ in \equ{tk} to the single-interface scattering matrix  in this case are logarithmic UV-divergent for \emph{any} correlation length $l_c>0$ . 

\begin{figure}
\includegraphics[scale=1.]{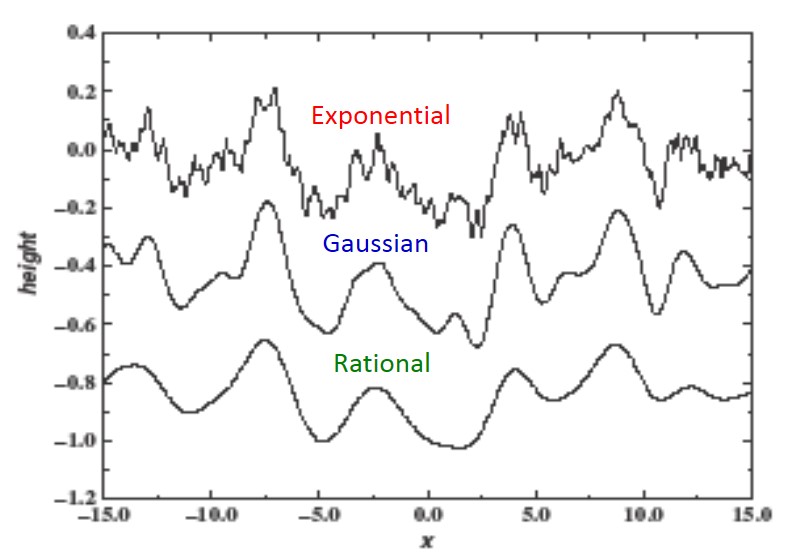}
\caption{Typical cross-sections of 2-dimensional profiles with different correlations  (reproduced from Ref.~\cite{Rough}).  From the top: profile with the exponential correlation $D_\text{Exp}(q)$ of \equ{ExpD}); profile with the Gaussian correlation $D_\text{Gauss}(q)$  of \equ{GaussD}; profile with a rational correlation (see text). The correlation length and variance are the same for all three profiles. For clarity the average height of the profiles differs by $-0.4$. Units are arbitrary. Note that only high-frequency components of the profiles differ significantly. } 
\label{profiles}
\end{figure} 

Experiment\cite{Palasantzas20051} does not distinguish Gaussian roughness correlations\footnote{$D_\text{Gauss}(q)=D_{\infty}(q)$ and $D_\text{Exp}(q)=D_{1/2}(q)$ in the class of $L_1$ correlations $\{D_s(q):=2\pi\sigma^2 l_c^2(1+\frac{q^2 lc^2}{2s})^{-s-1}$, with $s>0\}$.  The corresponding coordinate space correlation functions are $D_s(r)= \sigma^2\frac{2 (r\sqrt{2 s}/l_c)^s}{2^s\Gamma[s]} K_s( r\sqrt{2 s}/l_c)$. Ref.~\cite{Pala2008} uses a correlation in this affine class with $s=0.9$ for which the loop integral converges but is sensitive to contributions from high-momenta.}, and roughness profiles with correlation lengths $l_c\oP\ll 1$ are readily manufactured. Restricting the model to a particular form for the roughness correlation would not address the fact that the effective low-energy theory does not describe high-momentum contributions to loop integrals correctly. 

From a practical point of view the problem is that roughness corrections to the Casimir free energy and other low-energy observables are exceptionally \emph{sensitive} to high-frequency components of the profile because $\bG^|(k\sim
\infty, 0,0;\zeta)\sim k$ at large momenta. \fig{profiles} depicts typical roughness profiles to three different correlation functions with the same correlation length and variance: a) exponential as in \equ{ExpD}, b) Gaussian as in \equ{GaussD} and c) Rational as $D_\text{Rational}(r)=\sigma^2 /(1+(r/l_c)^2)^2$. It is evident from \fig{profiles} that the three profiles differ only in their high-frequency components. However, to leading order in the variance, corrections to the low-energy scattering matrix are extremely different for the three types of profiles.  The roughness correction diverges in the exponential case a) but is finite for profiles  b) and c). This sensitivity can be traced to the UV behavior of the 1-loop integrands  like that of \equ{txx0}. It is unphysical and an artifact of taking the low-energy effective theory beyond its limits.    

Analogous difficulties arise in any non-renormalizable low-energy effective field theory\cite{Leutwyler1994,Weinbergbook} and we here resort to a similar cure: whereas high momenta may dominate loop corrections to the scattering matrix, they generally are sufficiently suppressed in differences thereof. Differences of elements of the scattering matrix often can be reliably estimated within the framework of the low-energy effective field theory. However, phenomenological input is required to determine high-momentum contributions to loop integrals that are beyond the reach of the low-energy theory. 

One for instance can rewrite $t_{xx}^{(2)}(\bk,\zeta)$ of  \equ{tk} in the form,
\begin{align}\label{txxs}
t_{xx}^{(2)}(\bk,\zeta)&=t_{xx}^{(2)}(0,\zeta)+(t_{xx}^{(2)}(\bk,\zeta)-t_{xx}^{(2)}(0,\zeta))\\
&=t_{xx}^{(2)}(0,\zeta)-(\eps-1)^2\int \frac{d\bk'}{(2\pi)^2}\left( \frac{\kappa'\kappa'_\eps\cos^2\!\theta }{\kappa'_\eps+\eps\kappa'}+\frac{\zeta^2\sin^2\!\theta }{\kappa'_\eps+\kappa'}\right) (D(|\bk'-\bk|)-D(k'))\nonumber\\
&=t_{xx}^{(2)}(0,\zeta)-(\eps-1)^2\int \frac{d\bq}{(2\pi)^2}\left( \frac{\kappa'\kappa'_\eps (\hat \bk\cdot \hat \bk')^2 }{\kappa'_\eps+\eps\kappa'}+\frac{\zeta^2(1- (\hat \bk\cdot \hat \bk')^2) }{\kappa'_\eps+\kappa'}-\frac{\kappa\kappa_\eps (\hat \bk\cdot \hat \bq)^2 }{\kappa_\eps+\eps\kappa}-\frac{\zeta^2(1- (\hat \bk\cdot \hat \bq)^2) }{\kappa_\eps+\kappa}\right) D(q)\ ,\nonumber
\end{align}
where $\bq=\bk'-\bk, \kappa'_\eps=\sqrt{(\bk+\bq)^2+\zeta^2\eps(\zeta)}$ and  $\kappa_\eps=\sqrt{\bq^2+\zeta^2\eps(\zeta)}$ in the last expression. The one-loop  correction to $t_{xx}^{(2)}(0,\zeta)$ in\equ{txxs} converges for any $D(q)$ for which  
\bel{condD}
\vev{h^2(\bx)}=\int_0^\infty\frac{ q dq}{2\pi} D(q) = \sigma^2<\infty\ .
\ee 
More importantly, the correction to $t_{xx}^{(2)}(0,\zeta)$ in \equ{txxs} is of order $(k \sigma)^2$ and thus small at low transverse momenta for \emph{any} correlation length $l_c$ of the profile. This correction to $t_{xx}^{(2)}(0,\zeta)$ thus is reliably computed in the framework of the low-energy theory.

It remains to estimate $\bt^{(2)}(0,\zeta)$. This is the correction due to roughness to the (analytically continued) scattering matrix of an electromagnetic wave of frequency $\omega=i\zeta$ incident perpendicular to the rough plate. $\bt^{(2)}(0,\zeta)$ is a single-interface low-energy characteristic that, at least in principle, can be derived from ellipsometric measurements of the rough interface.  Instead of directly incorporating such experimental data, we here model the corrections of order $\sigma^2$ to the low-energy scattering matrix  by the coupling to surface plasmons induced by roughness. We determine the coupling by demanding that this phenomenological description of $\bt^{(2)}(0,\zeta)$ be consistent with the low-energy field theory in the limit of large correlation length and that the ideal metal limit exist at any correlation length.

\begin{figure}
\includegraphics[scale=0.5]{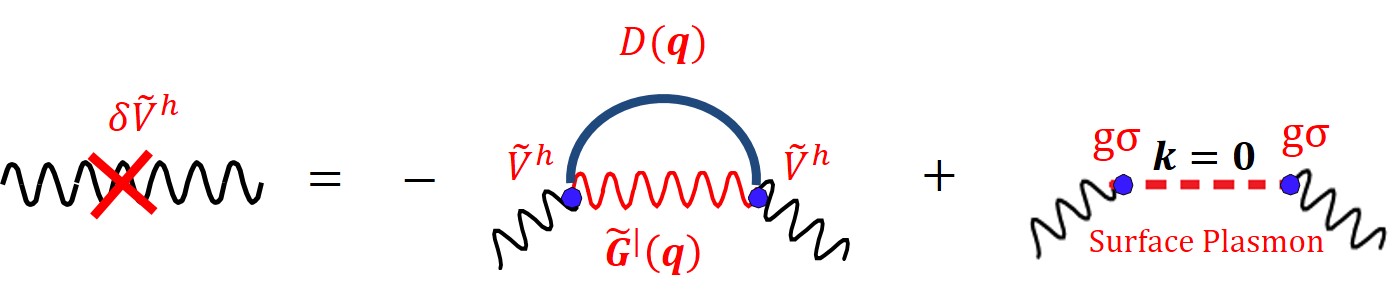}
\caption{The counter potential $\tilde \bV^h$ includes two contributions of order $\sigma^2$. It subtracts the one-loop contribution to the average scattering matrix at vanishing (transverse) momentum and replaces it by the phenomenological one. The latter is modeled by the tree-level plasmon contribution at vanishing transverse momentum. The plasmon couples to radiation due to the roughness of the surface  only and its coupling $g^2\sigma^2$ to this order is proportional to the variance of the roughness profile. The plasmon propagator (dashed) is the one-interface Green's function $\tilde\bG(z=z'=k=0)$. We show in the text that $g^2(\zeta/\oP,l_c\oP)=1$ at low frequencies.} 
\label{deltaVtilde}
\end{figure}

Roughness couples electromagnetic radiation to surface plasmons\cite{SPP}. At low transverse wave numbers this coupling is of the order of the rms-roughness $\sigma$. To order $\sigma^2$ the corresponding tree-level correction to the scattering matrix is schematically shown in \fig{deltaVtilde}. The diagram depicts the creation, propagation and subsequent annihilation of a surface plasmon by an incident electromagnetic wave.

For $\bk\rightarrow 0$ a surface plasmon on the interface of a flat plate at $z=0$ propagates with the dyadic,
\bel{plasmon}
\bG_\text{plasmon}(\bk\sim 0;\zeta)= \tilde\bG^|(k=0,z=z'=0;\zeta)\sim   \frac{\zeta}{1+\sqrt{\eps(\zeta)}}\text{diag}(1,1,0)\ .
\ee
To second order in $\sigma $, the correction $\bt^{(2)}(0,\zeta)$ to the scattering matrix at vanishing momentum transfer from surface plasmons thus is,
\bel{t0plas}
\bt^{(2)}(\bk=0,\zeta)\approx\bt_\text{Plasmon}^{(2)}(\bk=0,\zeta)=-\sigma^2 g^2\frac{\zeta  (\eps(\zeta)-1)^2}{ 1+\sqrt{\eps(\zeta)}}\text{diag}(1,1,0)\ ,  
\ee
where $g(\zeta/\oP; l_c\oP)$ is a dimensionless coupling that depends only on the frequency of the plane wave incident perpendicular to the rough plate. The coupling $g(\zeta/\oP; l_c\oP)$ in general is not calculable within this low-energy effective model and has to be determined phenomenologically. We argue below that $g^2\sim 1$ at low energies. 

Since $g(\zeta/\oP; l_c\oP)$ is a phenomenological \emph{function} rather than just a constant,  one could have directly modeled $\bt^{(2)}(\bk=0,\zeta)$.   However, the ansatz of \equ{t0plas} is consistent with the low-energy scattering theory  in the sense that roughness correlation functions for large correlation length $l_c\oP\gg1$ approach representations of the $\delta$-distribution\footnote{on the space of measurable  $L^0$ test-functions. The \emph{subtracted} loop integrand is in this class.},    
\bel{limlc}
\lim_{l_c\rightarrow\infty} D(\bq;l_c)=(2\pi)^2\sigma^2 \delta(\bq) 
\ee

Loop integrals in the limit $l_c\rightarrow\infty$ become trivial and furthermore involve only momenta $q\ll \oP$. Predictions of the low energy theory therefore should be reliable in the limit $l_c\rightarrow\infty$. Evaluating the loop integrals of \equ{tk} for $k\rightarrow 0$ using \equ{limlc} and comparing with the plasmon contribution of \equ{t0plas} this requires that ,
\bel{limg}
g(\zeta/\oP, l_c\oP\sim\infty)=1\ .
\ee
We will find that \equ{limg} not only ensures consistency, but also the existence of an ideal metal limit. It in addition ensures that the PFA to the Casimir free energy is recovered in the limit $ l_c\oP\rightarrow \infty$.

At finite $l_c \oP\lesssim 1$ the coupling $g(\zeta/\oP, l_c\oP\lesssim 1)$ in principle has to be determined phenomenologically. However, the coupling is severely constrained if we impose some theoretical requirements. Since the range of frequencies $\zeta$ that contribute to the Casimir energy satisfy $\zeta a\lesssim 1\ll \oP a$ and the plasmon coupling does not diverge at low frequencies, we in the following ignore the $\zeta$-dependence of $g(\zeta/\oP,l_c\oP)$ and for low frequencies approximate,
\bel{approxg}
g(\zeta/\oP,l_c\oP)\sim g(l_c\oP)\lesssim 1
\ee
in \equ{t0plas}. \equ{approxg} assumes that the plasmon coupling is strongest for an ideal metal $l_c\oP\gg 1$. Note that the fact that $g$ is dimensionless links the ideal metal to the  large $l_c$ limits. 

To order $\sigma^2$ the subtraction of the one-loop contribution $\bt^{(2)}(\bk=0,\zeta)$ and its replacement by phenomenological plasmon scattering is implemented by a (local in transverse coordinates) counter term potential $\delta \tilde\bV(\zeta,z)$ of the form,
\begin{align}\label{dtV}
\delta\tilde \bV^h&=\text{diag}(\delta V^h_{xx}(\zeta,z), \delta V^h_{yy}(\zeta,z), (\eps\theta(z)+\frac{1}{\eps}\theta(-z))\delta V^h_{zz}(\zeta,z), \ \text{with}\nonumber\\
\delta V^h_{xx}(\zeta,z)=\delta V^h_{yy}(\zeta,z)&=\delta(z) (\eps-1)^2\left[\frac{-g^2\sigma^2\zeta}{1 +\sqrt{\eps}}+\int_0^\infty\frac{ k dk}{4\pi}  D(k)\left(\frac{\kappa\kappa_\eps}{\eps\kappa+\kappa_\eps}+\frac{\zeta^2}{\kappa+\kappa_\eps}\right)\right]\nonumber\\
\delta V^h_{zz}(\zeta,z)&=-\delta(z)(\eps-1)^2\int_0^\infty\frac{ k dk}{2\pi}  D(k)\frac{ k^2}{(\eps\kappa+\kappa_\eps)}\ .
\end{align}
Note that the support of $\delta V^h(\zeta,z)$ is in the immediate vicinity of $z=0$ only.  Due to rotational and translational symmetry of the rough plate, this "counter potential" is local and diagonal but anisotropic\footnote{The product of distributions in $\delta\tilde \bV^h(\zeta,z)\propto\delta(z) (\eps(\zeta)\theta(z)+\eps^{-1}(\zeta)\theta(-z))$ here means that integration with a test function $f(z)\in L^0$ gives 
$ \int dz (\eps\theta(z)+\eps^{-1}\theta(-z))\delta(z) f(z):=\half(\eps\lim_{z\rightarrow 0_+}+ \eps^{-1}\lim_{z\rightarrow 0_-})f(z) $.}. 

As mentioned in Sec.~\ref{Roughness}, the counter potential may be interpreted as the modification of the dielectric permittivity (to order $\sigma^2$) in the vicinity of  the flat interface necessary to describe the rough interface with permittivity $\eps$ and roughness correlation $D_2(\bx-\by)$.  There is no compelling reason for perturbing about a flat interface with the \emph{same} permittivity as the rough one. We have seen that the expansion about a flat plate with the same permittivity is not consistent with the low-energy description, since it implies unacceptably high momenta in the loop integrals. Expanding instead about the bare permittivity function of\equ{Vslabs} yields a better controlled approximation and \equ{dtV} strongly suppresses high-momentum contributions to 1-loop. 

\section{Roughness Correction to the Casimir Free Energy of Order $\sigma^2$}
\label{secIV}
We now evaluate the roughness correction to the Casimir free energy within the framework of the improved low-energy effective field theory. From \equ{expandF} and \equ{bT2} we have altogether four contributions to order $\sigma^2$,
\bel{rough2F}
\Delta F_T^\text{Cas}[a]=\half\vev{\tr  \tilde\bV^h\bG^{|a|}}-\half\vev{\tr \tilde\bV^h\tilde \bG^|\tilde\bV^h \bG^{|a|}}+\half\tr  \delta\tilde\bV^h\bG^{|a|} -{\textstyle \frac{1}{4}}\vev{\tr \tilde \bV^h\bG^{|a|}\tilde\bV^h\bG^{|a|}}+\O(\sigma^3)\ .
\ee
We consider them in turn.
\begin{figure}
\includegraphics[scale=0.4]{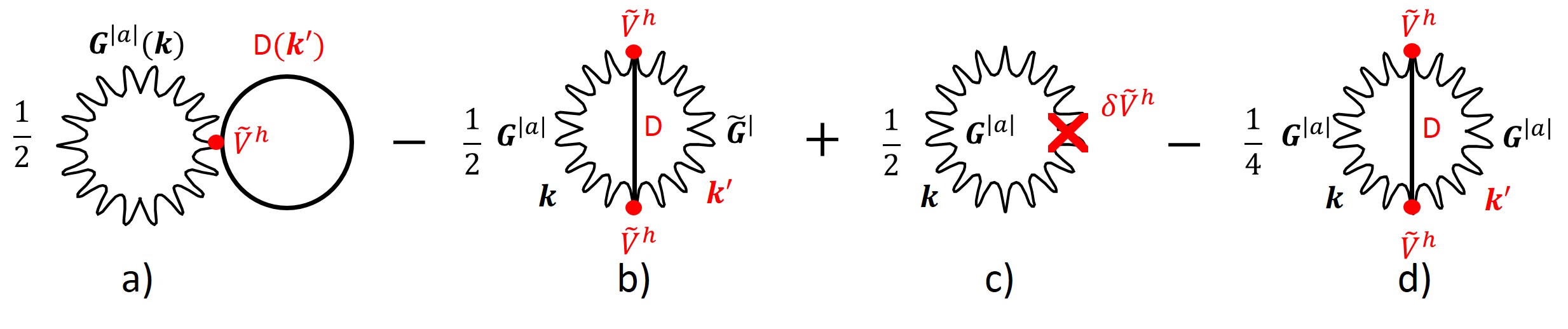}
\caption{Feynman diagrams for the contributions of order $\sigma^2$ to the roughness correction of the Casimir free energy of a rough and a flat interface. a) and b) give corrections from a single scattering off the rough surface and include only one factor of $\bG^{|a|}$. c) gives the contribution from the counter potential defined in \equ{dtV} whose two terms are shown in \fig{deltaVtilde}. This contribution eliminates the uncontrolled high-momentum contributions to the loop integral of b) in favor of a phenomenological (plasmon) description.  d) is the 2-scattering contribution of order $\sigma^2$ and includes two factors of $\bG^{|a|}$. The momenta in either loop of this term are  exponentially restricted to $k,k'\lesssim 1/(2 a)\ll \oP$ and no subtraction is required.  Wavy lines denote photon  propagators for a single flat interface, $\tilde\bG^|(\bk')$, or their correction, $\bG^|a|(\bk)$, due to the presence of a second flat interface at a mean distance $a$. Solid lines represent the Fourier transform $D(\bk-\bk')$ of  the roughness correlation function. A (red) dot indicates the effective anisotropic  interaction potential $\tilde V^h$ due to the roughness profile defined in \equ{tV}.  Combinatorical factors are shown but traces and momentum integrals have been suppressed.}
\label{DeltaF}
\end{figure}

\subsection{The Seagull Contribution $\half\vev{\tr  \tilde\bV^h\bG^{|a|}}$}
The first is the seagull contribution of \fig{DeltaF}a given by,
\begin{align}\label{seagull}  
\half\vev{\tr  \tilde\bV^h\bG^{|a|}}&=-\frac{A T}{2}\sum_n (\eps-1)\vev{ \int_0^\infty \frac{k dk}{2\pi}\int_0^{h(\bx)} \hspace{-1.6em}dz  (G_{xx}^{|a|}(k,z,z;\zeta)\!+\! G_{yy}^{|a|}(k,z,z;\zeta)\! +\!(\eps\theta(z)\!+\!\frac{\theta(-z)}{\eps})G_{zz}^{|a|}(k,z,z;\zeta))}\nonumber\\
&=-A T\sum_n \int_0^\infty \frac{k dk}{\pi} \kappa\kappa_\eps ( \frac{\bar r^2}{e^{2 a \kappa}-\bar r^2} +\frac{ r^2}{e^{2 a \kappa}- r^2} )\vev{\int_0^{h(\bx)}  \hspace{-1.6em}z dz }+\O(\sigma^3)\nonumber\\
&=-AT\sigma^2 \sum_{\zeta\in\{\zeta_n\}} \int_0^\infty \frac{k dk}{2\pi} \kappa\kappa_\eps ( \frac{\bar r^2}{e^{2 a \kappa}-\bar r^2} +\frac{ r^2}{e^{2 a \kappa}- r^2})+\O(\sigma^3) \ .
\end{align}
The expressions of \equ{Ga} in App.~\ref{appA} here have been  expanded for small  $z$. There are (as expected) no corrections of order $\sigma$ and the final line exhibits equally weighted contributions from both polarizations. Note that this remarkable simplification occurs only upon summation of all $\delta$-function contributions to $\bG^\parallel_{zz}$ - which gives an expansion in $\tilde \bV^h$ (defined in \equ{tV}), rather than in the original $\bV^h$.

This roughness contribution to the free energy is entirely local and does not depend on the correlation length $l_c$. The loop-integral over transverse momenta and the sum over Matsubara frequencies are exponentially restricted to momenta $2 a\kappa \lesssim 1$ and the evaluation of the seagull diagram using the low-energy propagators should be accurate for all $a\oP\gg 0.5$, that is for $a\gtrsim 12 \text{nm}$ in the case of gold plates. 
  
Due to the $\kappa_\eps$ factor of the integrand, the contribution of \equ{seagull}  is proportional to $\oP \sigma^2/a^4$ for $a\oP \gg 1\gg T a$ and diverges in the ideal metal limit. Fortunately the seagull is not the whole story to order $\sigma^2$.

\subsection{The Single Diffusive Scattering Contribution$\vev{\tr \tilde\bV^h\tilde \bG^|\tilde\bV^h \bG^{|a|}}$}
The other contribution to the Casimir free energy of order $\sigma^2$ from a single scattering off the rough interface corresponds to the diagram of \fig{DeltaF}b.   This unsubtracted  2-loop contribution is formally given by,
\begin{align}\label{fig2b} 
-\half\vev{\tr \tilde\bV^h\tilde \bG^|\tilde\bV^h \bG^{|a|}}=&-\frac{AT}{2}\sum_n\int \frac{d\bk d\bk'}{(2\pi)^4}  \tr\,\Big[D_{++}(q) \tilde\bG^{|(n)}_{++}(\bk')\bV^{(n)}_+ \bG_{++}^{|a|(n)}(\bk)\bV^{(n)}_++\\
&\hspace{-10em}D_{--}(q)\tilde \bG^{|(n)}_{--}(\bk')\bV^{(n)}_- \bG_{--}^{|a|(n)}(\bk)\bV^{(n)}_-+D_{-+}(q) \tilde\bG^{|(n)}_{-+}(\bk')\bV^{(n)}_+ \bG_{+-}^{|a|(n)}(\bk)\bV^{(n)}_-+D_{+-}(q)\tilde\bG^{|(n)}_{+-}(\bk')\bV^{(n)}_- \bG_{-+}^{|a|(n)}(\bk)\bV^{(n)}_+\Big],\nonumber
\end{align}   
with $q=|\bk-\bk'|$ and interaction vertices,
\bel{vertices}
\bV^{(n)}_+=(\eps(\zeta_n)-1)\text{diag}(1,1,\eps(\zeta_n)),\hspace{4em}\bV^{(n)}_-=(\eps(\zeta_n)-1)\text{diag}(1,1,1/\eps(\zeta_n))\ .
\ee
$\bG^{(n)}_{\pm\mp}(\bk):=\bG(\bk,z=0^\pm,z'=0^\mp;\zeta_n)$ denote one-sided limits of propagators. Explicit expressions are given in \equ{Gsnew}. The correlation functions $D_{\pm\mp}(q)$ of positive and negative components of the roughness profile are defined in \equ{Ds} and computed in App.~\ref{appB}.

A lengthy but otherwise straightforward evaluation of \equ{fig2b} using the expressions of \equ{Gsnew} and \equ{rotation} yields,
\begin{align}\label{fig2b1} 
&-\half\vev{\tr \tilde\bV^h\tilde \bG^|\tilde\bV^h \bG^{|a|}}=-\frac{AT}{2}\sum_n(\eps-1)^2\int_0^\infty\frac{k d k}{2\pi}\int_0^\infty\frac{ k' d k'}{(2\pi)^2} \int_{-\pi}^\pi d\theta D(\sqrt{k^2+k'^2-2 kk'\cos\theta})\times\\
&\times\Big[\frac{r (1-r^2) \zeta^2 }{2(e^{2 a \kappa}-r^2)\kappa_\eps}\left(\frac{\kappa'\kappa'_\eps\sin^2\!\theta}{\eps\kappa'+\kappa'_\eps}+\frac{\zeta^2\cos^2\!\theta}{\kappa'+\kappa'_\eps}\right)+\frac{ \bar r(1-\bar r^2)}{2 (e^{2 a\kappa}-\bar r^2)\eps}\left(\frac{\eps k^2 k'^2-\kappa^2_\eps \kappa'\kappa'_\eps \cos^2\!\theta}{\kappa_\eps(\eps\kappa'+\kappa'_\eps)}-kk'\bar r'\cos\theta- \frac{\kappa_\eps\zeta^2\sin^2\!\theta}{(\kappa'+\kappa'_\eps)} \right)\Big]\ .\nonumber
\end{align}
The signed correlation functions in \equ{fig2b} combine and \equ{fig2b1}  depends on the roughness correlation $D(|\bk-\bk'|)$ only.  In App.~\ref{appC} the integral over $\theta$ in \equ{fig2b1} is performed analytically for the class of correlations $D_s(q)$, but this angular integral in general has to be evaluated numerically.   More importantly, the leading term of order $\oP$ in the limit  $\oP\rightarrow\infty$ of \equ{fig2b1} cancels the leading asymptotic behavior $\propto\oP$ of the seagull term in \equ{seagull}. 

The limit of \equ{fig2b1} for large correlation length $l_c\gg 1/\oP$ is found  using  \equ{limlc} to trivially evaluate the $\bk'$-integrals. Some algebraic manipulations simplify the expression in this limit to,
\bel{fig2b2} 
-\half\vev{\tr \tilde\bV^h\tilde \bG^|\tilde\bV^h \bG^{|a|}}\xrightarrow{l_c\rightarrow\infty}AT\sigma^2\sum_n\int_0^\infty\frac{k d k}{2\pi}\kappa(\kappa_\eps-\kappa)\left(\frac{r^2}{e^{2 a \kappa}-r^2}+\frac{ \bar r^2}{ e^{2 a\kappa}-\bar r^2}\right)\ .
\ee

\subsection{The Counterterm Correction}
As for $\bt^{(2)}$ in \equ{tk}, the loop-integral of\equ{fig2b1} generally includes high momentum contributions $k'\gg\oP$ for which the low-energy description is not justified. The same 1-loop counter potential of \equ{dtV} that corrects roughness corrections to the scattering matrix to 1-loop also removes the uncontrolled high-momentum contributions to the Casimir free energy  and replaced them by the phenomenological plasmon contribution.     
 
The correction of the Casimir free energy by this counter potential is shown diagrammatically in \fig{DeltaF}c and the two Feynman diagrams of this counter term are depicted in \fig{deltaVtilde}. To order $\sigma^2$ the contribution to the Casimir free energy from the counter potential  $\delta\tilde\bV$ of \equ{dtV} is, 
\begin{align}\label{fig2c}
\half\tr \delta\tilde\bV \bG^{|a|}&=\frac{AT}{2}\sum_n(\eps-1)^2\int_0^\infty\frac{k d k}{2\pi}\int_0^\infty\frac{ k' d k'}{2\pi}  D(k')\Big[\frac{ \bar r(1-\bar r^2) k^2}{2 (e^{2 a\kappa}-\bar r^2)\kappa_\eps}\frac{  k'^2}{(\eps\kappa'+\kappa'_\eps)}+\\
&\hspace{10em}\left(\frac{r (1-r^2) \zeta^2 }{2(e^{2 a \kappa}-r^2)\kappa_\eps}-\frac{ \bar r(1-\bar r^2)\kappa_\eps}{2(e^{2 a\kappa}-\bar r^2)\eps}\right)\left(\frac{\kappa'\kappa'_\eps/2}{\eps\kappa'+\kappa'_\eps}+\frac{\zeta^2/2}{\kappa'+\kappa'_\eps}-\frac{g^2 \zeta}{1 +\sqrt{\eps}}\right)\Big]\nonumber
\end{align}
This correction to the Casimir free energy remains finite in the ideal metal limit when \equ{limg} is satisfied. The existence of this limit is assured by the consistency of the low-energy theory in the limit $l_c\gg 1/\oP$. Using \equ{limlc},  the counterterm correction of \equ{fig2c} for $l_c\gg\oP$ becomes,
\bel{fig2c2} 
\half\tr \delta\tilde\bV \bG^{|a|}\xrightarrow{l_c\rightarrow\infty}AT\sigma^2\sum_n(g^2-1)\zeta(\sqrt{\eps}-1)\int_0^\infty\frac{k d k}{2\pi}\kappa\left(\frac{ r^2}{e^{2 a \kappa}-r^2}+\frac{\bar r^2\kappa_\eps^2}{(e^{2 a\kappa}-\bar r^2)(\eps k^2+\kappa_\eps^2 )}\right)\xrightarrow{g^2\rightarrow 1}0\ ,
\ee
and vanishes when \equ{limg} is enforced. This should be expected of a model that is valid at low energies. Note that the reason magnetic and electric modes do not enter the counter term correction symmetrically even at large correlation length is because we subtracted at $\bk=0$: the factor $\kappa_\eps^2/(\eps k^2 +\kappa_\eps^2)$ in \equ{fig2c2} differs from unity in order $k^2/\oP^2$ only. 

\subsection{Contributions of Second Order in the Roughness Scattering Matrix}
Both loop integrals of this contribution (represented in \fig{DeltaF}d) to the Casimir free energy are exponentially constrained  to low momenta $k,k'\lesssim 1/(2 a)\ll \oP$ -- a regime in which the low-energy description is expected to hold. We find that,
\begin{align}\label{fig2d} 
&-{\textstyle \frac{1}{4}}\vev{\tr \tilde\bV^h\tilde \bG^{|a|}\tilde\bV^h \bG^{|a|}}=-\frac{AT}{16}\sum_n(\eps-1)^2\int_0^\infty\frac{k d k}{2\pi}\int_0^\infty\frac{ k' d k'}{(2\pi)^2} \int_{-\pi}^\pi d\theta D(\sqrt{k^2+k'^2-2 kk'\cos\theta})\times\\
&\times\Big[\frac{r (1-r^2) \zeta^2 }{(e^{2 a \kappa}-r^2)\kappa_\eps}\left(\frac{r' (1-r'^2) \zeta^2\cos^2\!\theta }{(e^{2 a \kappa'}-r'^2)\kappa'_\eps}-\frac{2\bar r'(1-\bar r'^2)\kappa'_\eps\sin^2\!\theta}{(e^{2a\kappa'}-\bar r'^2)\eps}\right)+\frac{ \bar r\bar r'(1-\bar r^2)(1-\bar r'^2)}{ (e^{2 a\kappa}-\bar r^2)(e^{2 a\kappa'}-\bar r'^2)}\left(\frac{k^2 k'^2}{ \kappa_\eps\kappa'_\eps}+\frac{2 k k'\cos\theta}{\eps }+\frac{\kappa_\eps\kappa'_\eps\cos^2\!\theta}{\eps^2 }\right) \Big]\ .\nonumber
\end{align}    

For profiles with large correlation length $l_c\gg 2a \gtrsim 1/\oP$ \equ{fig2d} simplifies to
\bel{fig2d2}
-{\textstyle \frac{1}{4}}\vev{\tr \tilde\bV^h\tilde \bG^{|a|}\tilde\bV^h \bG^{|a|}}\xrightarrow{l_c\rightarrow\infty} -AT\sigma^2\sum_n\int_0^\infty\frac{k d k}{2\pi} \kappa^2\left(\frac{r^4}{(e^{2 a \kappa}-r^2)^2}+\frac{\bar r'^4}{(e^{2a\kappa'}-\bar r'^2)^2}\right), 
\ee
when \equ{limlc} holds.
\subsection{The Limit $l_c\gg \text{max}(1/\oP,a)$: the Proximity Force Approximation}
Although $l_c\gg a$ is a necessary condition for the PFA, the limiting expressions of Eqs.~(\ref{fig2b2})~and~(\ref{fig2c2}) evidently hold only when $l_c$ is large compared to $a$ \emph{and} $1/\oP$ . The latter restriction arises because the scattering matrix locally can be approximated by a flat surface only if the plasma length is shorter than the typical length scale of the surface structure.

For a rough profile with $l_c\gg \text{max}(1/\oP, a)$ Eqs.~(\ref{fig2b2}),~(\ref{fig2c2})~and~(\ref{fig2d2}) should all be reasonable approximations. Including the seagull term of \equ{seagull}, the roughness correction to the Casimir free energy of\equ{rough2F} in the limit of large correlation length  $l_c\gg \text{max}(1/\oP, a)$ is,
\begin{align}\label{PFApprox}
\Delta F_T^{Cas}[a]&  \xrightarrow{l_c\rightarrow\infty}-AT\sigma^2\sum_n\int_0^\infty\frac{k d k}{2\pi}\kappa^2\left(\frac{r^4}{(e^{2 a \kappa}-r^2)^2}+\frac{r^2}{e^{2 a \kappa}-r^2}+\frac{\bar r'^4}{(e^{2a\kappa'}-\bar r'^2)^2}+\frac{ \bar r^2}{ e^{2 a\kappa}-\bar r^2}\right)\nonumber\\
&=\half\sigma^2 \frac{\partial^2}{\partial a^2}\frac{AT}{2}\sum_n\int_0^\infty\frac{k d k}{2\pi}\ln\left(1-r^2 e^{-2 a \kappa}\right)+\ln\left(1-\bar r^2 e^{-2 a \kappa}\right)=\half\sigma^2 \frac{\partial^2}{\partial a^2} F_T^\parallel(a)\ , 
\end{align}  
where $F_T^\parallel(a)$ is the Casimir free energy for two flat parallel semi-infinite slabs at a separation $a$ given by \equ{CasF}. This is precisely the roughness correction in PFA for a rough surface with $\vev{h(\bx)}=0$ and $\vev{h^2(\bx)}=\sigma^2$.  Although trivial, one should note that the PFA here emerges in the limit of large $l_c$ from requiring consistency of the low-energy effective field theory. It is due to the absence of high-momentum contributions in this limit and does not require any additional phenomenological input. 

\subsection{Ideal Metal Limit $\eps\rightarrow\infty$}
 It perhaps is remarkable that the requirement of \equ{limg} not only guarantees that the PFA is recovered in the $l_c\rightarrow\infty$ limit but also ensures the existence of an ideal metal limit. If $g^2$ is analytic at $zeta=0$ one can argue  that $\zeta/\oP$ and $1/(l_c\oP)$ (see \equ{limits}) corrections are absent and $g^2$ for large $\oP$ has the expansion $g^2= 1+\O( \zeta^2/\oP^2)$. The ideal metal limit in this case is uniquely given by,
\begin{subequations}
\label{idealmetal}
\begin{align}
\half\vev{\tr  \tilde\bV^h\bG^{|a|}}-\half\vev{\tr \tilde\bV^h\tilde \bG^|\tilde\bV^h \bG^{|a|}}&=-AT\sum_n\int_0^\infty\frac{k d k}{2\pi}\int_0^\infty\frac{ k' d k'}{(2\pi)^2} \int_{-\pi}^\pi d\theta D(\sqrt{k^2+k'^2-2 kk'\cos\theta})\times\nonumber\\
&\hspace{5em}\times\Big[ \frac{(\zeta^2+kk'\cos\!\theta)^2+\kappa^2\kappa'^2}{\kappa\kappa'(e^{2 a \kappa}-1)}-\frac{4k\zeta^2(k-k'\cos\!\theta)e^{2\kappa a}}{\kappa^2(e^{2 a \kappa}-1)^2}\Big]\label{firstideal}\\
-{\textstyle \frac{1}{4}}\vev{\tr \tilde\bV^h\tilde \bG^{|a|}\tilde\bV^h \bG^{|a|}}&=-AT\sum_n\int_0^\infty\frac{k d k}{2\pi}\int_0^\infty\frac{ k' d k'}{(2\pi)^2} \int_{-\pi}^\pi d\theta D(\sqrt{k^2+k'^2-2 kk'\cos\theta})\times\nonumber\\
&\hspace{5em}\times\frac{(\zeta^2+kk'\cos\!\theta)^2+\kappa^2\kappa'^2}{(e^{2 a \kappa}-1)(e^{2 a \kappa'}-1)\kappa\kappa'}\label{secondideal}\\
\half\tr \delta\tilde\bV \bG^{|a|}&=AT\sum_n\int_0^\infty\frac{k d k}{2\pi}\int_0^\infty\frac{ k' d k'}{2\pi}  D(k')\Big[\frac{2k^2k'^2+(\kappa^2+\zeta^2)(\kappa'-\zeta)^2}{2(e^{2a\kappa}-1)\kappa\kappa'}\Big]\label{thirdideal}\ .
\end{align} 
\end{subequations}
Note that the counter term contribution of \equ{thirdideal} does not vanish and cancels the contribution from high $k'$-momenta in\equ{firstideal} also for the ideal metal. High-momentum contributions to the roughness correction thus persist in the ideal metal limit. Without counter term this perturbative correction would diverge for $l_c\rightarrow 0$ (and for some correlations would diverge for \emph{all} $l_c$). This apparently is at odds with exact calculations for square-wave profiles\cite{Emig2004} and demands an explanation. The reason for convergence of the exact calculations in the limit $l_c\rightarrow 0$ (and divergence of the unsubtracted perturbation theory) for such profiles is subtle and related to the fact that for $l_c\ll \sigma$ the leading term in the exact calculation is $\O(\sigma)$ and not $\O(\sigma^2)$ as perturbation theory suggests\cite{Emig2004}. The non-analytic dependence on $\sigma$ for $l_c\rightarrow 0$ arises due to an effective UV-cutoff in the exact calculation  of $\O(\sigma)$ -- there is no other scale to compare with in this limit. Ignoring this effective cutoff (as a perturbative expansion in $\sigma$ does ) leads to an UV-divergent expression in the limit $l_c\rightarrow 0$.  The non-analyticity of the exact result  for $\sigma/a\ll1$ in the limit $l_c\rightarrow 0$ is only possible if wave-numbers of order $1/\sigma$ of the profile contribute significantly. The non-analyticity in $\sigma$ in this sense implies that high-momenta $1/a<k'<1/\sigma$ must dominate the exact Casimir energy calculation for an ideal metal in the limit  $0\leq l_c<\sigma\sim 0$.  A simple model that qualitatively reproduces this explanation of the non-analytic dependence on $\sigma$ is obtained by replacing $l_c\rightarrow l_c+\gamma \sigma$ in the Gaussian correlation function of \equ{GaussD} where the constant $\gamma$ is of $\O(1)$. For $l_c\gg\sigma$ one recovers the quadratic perturbative dependence on $\sigma$ in leading order, but for $0\leq l_c\ll\sigma\rightarrow 0$ the $k'$ integral of \equ{firstideal}  is of order $\sigma^2/(l_c+\gamma\sigma)\xrightarrow{l_c\ll \sigma} \sigma/\gamma$ as in the exact calculation. The UV-divergence $\propto 1/\sigma^3$ of the $k'$-integral that gives this leading (non-analytic) behavior is due to momenta $k'\sim 1/\sigma\gg 1/a$.  Although the exact evaluation of such high-momentum contributions is of itself correct, the low-energy description used to compute them is not justified.   The fact that the plasmon contributes and the counter term of\equ{thirdideal} removes high momentum contributions even for an ideal metal  indirectly supports the assertion that roughness corrections of real materials in fact remain analytic in the variance $\sigma^2$ even in the limit of uncorrelated roughness.         

\subsection{The Limit of Uncorrelated Roughness and the Plasmon Coupling $g^2$} 
The high-roughness limit $l_c\ll 1/\oP$ is obtained by examining the loop integrals in Eqs.~(\ref{seagull}),~(\ref{fig2b1}),~and~(\ref{fig2c}) at large momentum transfers $q=|\bk'-\bk|$. In the limit of uncorrelated roughness $l_c\rightarrow 0$ the correction is,
\begin{align}\label{lc0}  
\Delta F_T^\text{Cas}[a]&\xrightarrow{l_c\rightarrow 0} -AT\sigma^2 \sum_n \int_0^\infty \frac{k dk}{2\pi} \left[ \frac{\bar r^2\kappa\kappa_\eps(2\frac{\eps-1}{\eps+1} k^2 +\kappa^2_\eps-g^2(\sqrt{\eps}-1)\zeta\kappa_\eps)}{(e^{2 a \kappa}-\bar r^2)(k^2\eps+\kappa^2_\eps)} +\frac{ r^2\kappa(\kappa_\eps-g^2(\sqrt{\eps}-1)\zeta)}{e^{2 a \kappa}- r^2}\right]\ .
\end{align}
Note that the correction to the Casimir free energy for $l_c=0$  is strictly negative when $g^2\leq 1$. The Casimir free energy of a rough interface thus is always larger in magnitude than of a flat one at the same average separation. We believe this is the result of two opposing effects. The specular reflection off a rough surface with vanishing $l_c$ but finite $\sigma$ never is quite the same as that off a flat interface with the \emph{same} bulk permittivity: the situation  is analogous to the change in bulk permittivity due to the inclusion of sub-wavelength spheres of a different material. Since the included "material" in this case is vacuum with $\eps=1$, the effective reflection coefficient decreases compared to that for the flat plate. This effect by itself would tend to decrease  the Casimir free energy in magnitude for $l_c\rightarrow 0$. However, this decrease is more than compensated by the reduced separation to this effective interface.

The ideal metal limit of \equ{lc0} exists only for $g^2\rightarrow 1$ and is analytically given by,
\bel{lc0metal}
\Delta F_T^\text{Cas}[a, l_c\ll1/\oP\rightarrow 0]=-AT\sigma^2 \sum_n \int_0^\infty \frac{k dk}{2\pi}\frac{\zeta (\kappa^2+\zeta^2)}{\kappa(e^{2 a \kappa}-1)}\xrightarrow{T\rightarrow 0}-\frac{9A\sigma^2}{32\pi^2 a^5}\zeta(5)\approx -0.02955 \frac{A\sigma^2}{a^5}\ .
\ee
The ideal metal and $l_c\rightarrow 0$ limits in fact commute and  $g^2\rightarrow 1$ is required for the ideal metal limit to exist. Assuming that $g^2(\zeta l_c, l_c\oP)$ is analytic in both arguments, the existence of an ideal metal limit implies,  
\bel{limits}
1=\lim_{\genfrac{}{}{0pt}{1}{\oP\rightarrow\infty}{l_c\oP=\beta}} g^2(\zeta l_c, l_c\oP)=g^2(0, \beta)\ . 
\ee     
We therefore have that $g^2=1$ at low frequencies for any value of $l_c$ and $\oP$. We in the following therefore consider only,
\bel{g1}
g^2=1\ .
\ee  
\section{The Effective Low-Energy Field Theory of Electromagnetic Interactions with Rough Surfaces.}
\label{secV}
Although we obtained a roughness correction that is compatible with the low-energy theory of Schwinger by a Greens-function approach, it is instructive to construct the effective low energy field theory from which these corrections derive. The effective field theory allows one to in principle explore other approximations and corrections. It also provides a general framework for systematically taking into account higher orders or for including other interactions.  In this formulation the necessity of counter terms furthermore is readily apparent.

\subsection{The Generating Functional of Roughness Correlations}
The construction of the field theory is based on the generating function of the roughness correlation functions rather than the roughness correlations themselves. This approach was already used in the scalar case\cite{HuaMartinin2012}. The  $n$-point roughness correlation functions for an interface of (large) area $A$ with a \emph{particular} profile $h(\bx)$ are the averages,
\begin{align}\label{def-D}
D_1&=\vev{h(\bx_1)}&:=&A^{-1}\int_A  h(\bx+\bx_1) d\bx\nonumber\\
D_2(\bx_1-\bx_2)&=\vev{h(\bx_1)h(\bx_2)}&:=&A^{-1}\int_A  h(\bx+\bx_1)h(\bx+\bx_2) d\bx\nonumber\\
&\vdots &\vdots& \\
D_n(\bx_1-\bx_2,\dots,\bx_{n-1}-\bx_n)&=\vev{h(\bx_1)\dots h(\bx_n)}&:=&A^{-1}\int_A  h(\bx+\bx_1)\dots h(\bx+\bx_n) d\bx\ .\nonumber
\end{align}
The interface is assumed large enough for boundary effects to be negligible. Transverse translational invariance then implies that these correlations depend only on \emph{differences} of the transverse coordinates\footnote{For exact translational invariance, the finite parallel flat surfaces could be replaced by two concentric two-dimensional tori of area $A$.}. Isotropy of the roughness profile yields further restrictions:    the $n$-point correlation function in this case depend only on  \emph{distances} between the  points. We assume that the profile and therefore all $n$-point correlation functions of\equ{def-D} can, at least in principle, be measured when the rough interface is far removed from the other. The mean separation $a$ between the two interfaces is determined so that \equ{vev0} holds, that is the (constant) one-point function $D_1$ vanishes. We formally collect all roughness correlation functions of\equ{def-D} for a particular profile $h(\bx)$ in a single generating functional $Z_h[\alpha]$,
\bel{Zh}
Z_h[\alpha]=\sum_{n=2}^\infty \frac{1}{n!}\doubleint \alpha(\bx_1)\alpha(\bx_2)\dots\alpha(\bx_n)D_n(\bx_1,\dots,\bx_n) d\bx_1d\bx_2\dots d\bx_n\ .
\ee
Note that $Z_h$ depends on a \emph{particular} profile $h(\bx)$. None of the $D_n$ are averages over different profiles. For another profile some or all of the correlations defined by \equ{def-D} change and so does the functional $Z_h[\alpha]$.  

However, for constructing the field theory it is expedient to directly model $Z_h[\alpha]$ instead of computing the individual correlation functions of a given profile $h(\bx)$.  With the restriction of \equ{vev0} that the 1-point function vanishes, the simplest model of a rough interface is entirely determined by its $2$-point correlation $D_2$. The generating functional of such a (quadratic) Gaussian model is of the form,
\bel{ZhG}
Z^{G}_h[\alpha]=\exp \half(\alpha|D_2|\alpha)\ ,
\ee
with,
\bel{scalar2d}
(\alpha|D_2|\alpha):=\doubleint \alpha(\bx)D_2(\bx-\by)\alpha(\by)d\bx d\by \ .
\ee
In general \equ{ZhG} is only the leading quadratic term in a cumulant expansion of $Z_h$. A Gaussian generating functional relates all higher order correlations to the 2-point function.  In App.~\ref{appB} we for instance determine signed correlation functions in terms of $D_2$ for such a model. Stochastic roughness is fully described by the covariance of the profile and a Gaussian model by definition is exact in this case. A Gaussian model for the generating functional also suffices to obtain corrections to the free energy and the scattering matrix to leading order in the variance of the roughness profile.  To order $\sigma^2$ the correlations of a periodic 1-dimensional profile $h^\o(\bx)=\sigma\sin(\o x)$ can be found using a Gaussian model, but the four-point correlation function in this case is only half of what the Gaussian model asserts,
\begin{align}
\label{periodic}
D_2^\o(\bx-\by)&=\frac{\sigma^2}{2}\cos(\o(x-y))\ \text{  but  },\\ D_4^\o(\bx_1,\bx_2,\bx_3,\bx_4)&=\half (D_2^\o(\bx_1-\bx_2)D_2^\o(\bx_3-\bx_4)+\nonumber\\
&\ \ \ \ +D_2^\o(\bx_1-\bx_3)D_2^\o(\bx_2-\bx_4)+ D_2^\o(\bx_1-\bx_4)D_2^\o(\bx_2-\bx_3))\ .\nonumber
\end{align}
To correctly obtain effects due to a periodic profile to order $\sigma^4$ requires the inclusion of a $4^\text{th}$ order cumulant. Note that the 2-point correlation $D_2^\o$ in\equ{periodic} of a periodic corrugated profile is not positive definite and has no probabilistic interpretation. However, in momentum space it is proportional to the sum of two $\delta$-functions and therefore positive semi-definite. 

The  basis for a field theoretic approach to roughness is that \emph{any} analytic functional $R[h]$ of the profile $h(\bx)$ with \emph{translation invariant} coefficients can be evaluated using $Z_h[\alpha]$. To show this, consider a typical monomial in the Taylor expansion of $R[h]$ for small profiles,
\begin{align}
\label{evalf}
\doubleint & d\bx_1d\bx_2\dots d\bx_n R_n(\bx_1-\bx_2,\dots,\bx_{n-1}-\bx_n) h(\bx_1)h(\bx_2)\dots h(\bx_n) =\nonumber\\
&=\frac{1}{A}\int_A d\bx \doubleint d\bx_1d\bx_2\dots d\bx_n R_n(\bx_1-\bx_2,\dots,\bx_{n-1}-\bx_n) h(\bx+\bx_1)h(\bx+\bx_2)\dots h(\bx+\bx_n) \nonumber\\
&=\doubleint d\bx_1d\bx_2\dots d\bx_n R_n(\bx_1-\bx_2,\dots,\bx_{n-1}-\bx_n) D_n(\bx_1-\bx_2,\dots,\bx_{n-1}-\bx_n)\\
&=\left. \doubleint d\bx_1d\bx_2\dots d\bx_n R_n(\bx_1-\bx_2,\dots,\bx_{n-1}-\bx_n) \frac{\delta}{\delta\alpha(\bx_1)}\frac{\delta}{\delta\alpha(\bx_2)}\dots\frac{\delta}{\delta\alpha(\bx_n)} Z_h[\alpha]\right|_{\alpha=0}\ .\nonumber
\end{align}
The first equality in\equ{evalf} is due to the  translational invariance of the coefficient functions $R_n$  [but it does not assume any regularity of the profile $h(\bx)$ itself]. No further assumptions are required to show that \equ{evalf} holds for \emph{any} profile of a sufficiently large interface. The second equality in\equ{evalf} implies that the result is proportional to the area $A$.  Assuming that all coefficient functions $R_n$ in the Taylor expansion of $R[h]$ are translation invariant and that the expansion converges for the particular profile,\equ{evalf} implies that one may evaluate $R[h]$ for any \emph{particular} profile $h(\bx)$ by applying the  corresponding functional derivative operator on $Z_h[\alpha]$,
\bel{evalF}
R[h]=R[\frac{\delta}{\delta \alpha}] \;Z_h[\alpha]\Big|_{\alpha=0}\ .
\ee
\subsection{The Partition Function of the Low-Energy Effective Field Theory}
In the presence of external sources of polarization $\bP_n(\bx,z)=\bP(\bx,z;\zeta_n)$, Schwinger's free energy for two parallel interfaces is given by \equ{Fn0}.  The partition function for a flat and a rough interface described by the profile $h(\bx)$  corresponding to the potential $\bV(\zeta,h(\bx),z)$ of \equ{defbV} therefore formally is, 
\bel{genfunction}
Z_T[\bP,h]=\exp\Big[-\frac{1}{T}(F_T^\parallel(a)+\delta F[h])\Big]\prod_n \exp\Big[-\frac{1}{T}({\cal V}_n[h]+\delta{\cal V}^h_n)\Big]\exp\Big{[}\frac{T}{2}\{ \bP_n\vert \bG^{\parallel(n)}\vert  \bP_n\}\Big{]}\ ,
\ee
where ${\cal V}_n[h]$ is the functional derivative operator,
\bel{intV}
{\cal V}_n[h]=-\frac{1}{2}\int  d\bx\int_0^{h(\bx)} \hspace{-1.3em}dz\;  \frac{\delta}{\delta\bP_n(\bx,z)}\cdot(\beps(\zeta_n)-\one)\cdot\frac{\delta}{\delta\bP^\dagger_n(\bx,z)}
\ee
representing the interaction of the $n$-th Matsubara mode with the roughness profile $h(\bx)$. 
\subsection{Counter Terms of the Low-Energy Effective Field Theory}
The counter potential of \equ{dtV} corresponds to a functional derivative operator of the form,
\bel{deltaV}
\delta{\cal V}^h_n=\frac{1}{2}\int dz\int  d\bx \frac{\delta}{\delta\bP_n(\bx,z)}\cdot \delta\bV^h(\zeta_n,z)\cdot\int  d\by \frac{\delta}{\delta\bP^\dagger_n(\by,z)}\ .
\ee
It corrects for polarization effects due to surface roughness. Note that the counter potential of \equ{dtV} in \equ{deltaV} has support in the immediate vicinity of the plane at $z=0$ only and does not depend on the transverse position $\bx$ nor on the mean separation $a$ of the two interfaces.  The counter potential $\delta\bV^h(\zeta)$ should ensure that the scattering of electromagnetic waves incident perpendicular to the rough surface is reproduced. 

We in addition have to include a counterterm $\delta F[h] $ to the free energy that is a functional of the profile $h(\bx)$. It vanishes for $h(\bx)=0$ and has the expansion,    
\bel{Vc}
\delta F[h]=c_0+\int d\bx\ h(\bx)c_1(a,T)+\frac{1}{2}\doubleint d\bx d\by\ c_2(\bx-\by)h(\bx)h(\by)+\frac{1}{6}\int\hspace{-.9em}\int\hspace{-.9em}\int d\bx d\by d\bm{z}\ c_3(\bx-\bm{z},\by-\bm{z})h(\bx)h(\by)h(\bm{z})+\dots
\ee 
with translation-invariant $n$-point coefficient functions $c_n$ that depend only on transverse coordinate differences. These coefficient functions are used to systematically remove corrections to the correlation functions of the profile $h(x)$ in the presence of electromagnetic interactions.  The constant 1-point counter term $c_1(a, T)$ ensures that $\vev{h(x)}=0$ at any separation $a$ and temperature $T$. $c_1(a, T)$ is the only coefficient that may depend on $a$ and $T$ because its contribution to the free energy in fact vanishes for profiles that satisfy \equ{vev0}.   The higher order terms of $\delta F[h]$ are constructed so that connected correlation functions of the profile at $T=0$ are the prescribed ones when the second flat interface is removed. They do not depend on the temperature $T$ nor on the separation $a$. This ensures that,
\bel{condition}
\frac{\partial}{\partial T}\delta F[h]=\frac{\partial}{\partial a}\delta F[h]=0 \ \text{for any profile for which} \int_A d\bx h(\bx)=0 \ .
\ee
This counter-term to the free energy therefore does not affect  thermodynamic state functions like the enthropy or pressure. It cancels  loop contributions to the energy (at $T\rightarrow 0$) when the flat interface is removed ($a\rightarrow \infty$). The Casimir free energy remains (its finite, $a$-dependent value at $T=0$ is the Casimir energy).  
   
\begin{figure}
\includegraphics[scale=0.5]{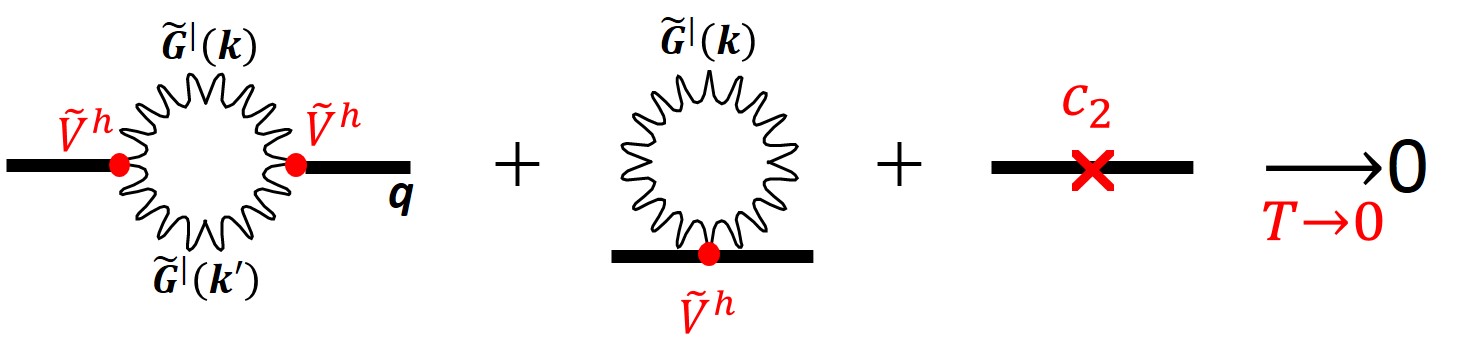}
\caption{One-loop Feynman diagrams for the counter term $c_2(q)$. $c_2(q)$ is determined by demanding that the (prescribed) 2-point roughness correlation of a single plate at $T=0$ is not corrected. We here consider 1-loop contributions only. } 
\label{c2counter}
\end{figure}

In obtaining  the Casimir free energy by the Green's function method the contribution to the free energy from the counter term coefficient $c_2(\bx-\by)$ was implicitly taken into account by subtracting $\Delta F_T[h,\infty]$ in\equ{FC}.  Requiring the absence of one-loop corrections to the 2-point roughness correlation  at large separation $a$  and temperature $T=0$ determines $c_2(\bq)$. The Feynman diagrams involved in this condition are shown in \fig{c2counter}. The counter term $c_2$ also ensures that there is no single-interface correction to the Casimir energy at $T=0$. For $T>0$ a finite $a$-independent contribution to the single-interface free energy remains that we have not calculated here. 

\begin{figure}
\includegraphics[scale=0.32]{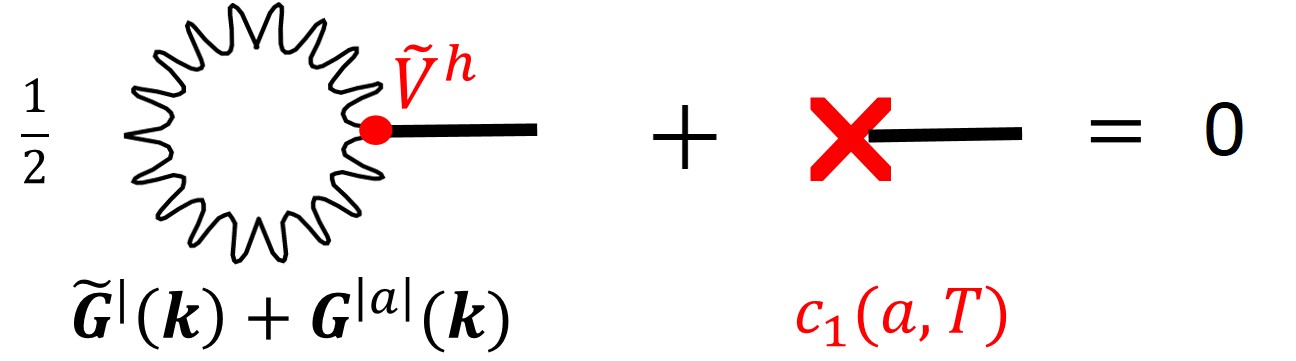}
\caption{Cancellation of tadpoles by the counter term $c_1(a,T)$ at one loop. Summation to all orders of the $\delta$-function contribution to $G_{zz}^\parallel$ replaces $\bG^\parallel$ by $\tilde\bG^|+\bG^{|a|}$ and $\bV^h$ by $\tilde \bV^h$. } 
\label{tadpoles}
\end{figure}

\begin{figure}
\includegraphics[scale=0.3]{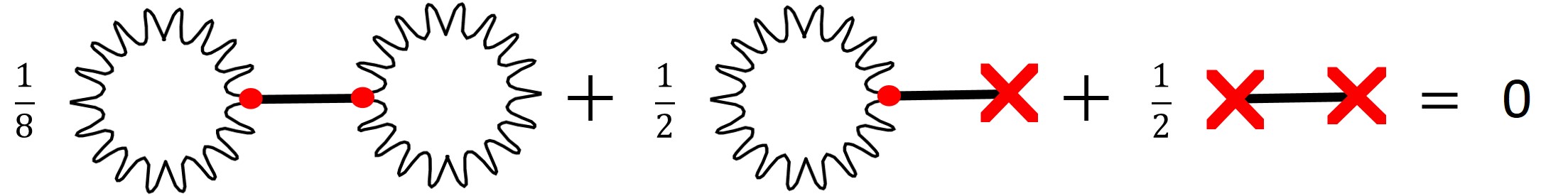}
\caption{1-particle reducible dumbbell contributions to the free energy that are cancelled by the $c_1$ counter term given in\equ{defc1}. 1-particle reducible contributions to the free energy are of order $1/T$ at low temperatures and would violate Nernst's theorem.} 
\label{dumbbells}
\end{figure}

The Green's function approach implicitly also accounted for contributions of $c_1(a,T)$ by simply assuming  that \equ{vev0} holds to order $\sigma^2$.  $c_1(a, T)$ cancels tadpole contributions to the scattering matrix (see \fig{tadpoles}) and 1-particle reducible contributions to the Casimir free energy like those of\fig{dumbbells} vanish in this case.   

We defined the mean separation $a$ by \equ{vev0} and demanding that corrections to $\vev{h_\pm(\bx)}$ vanish determines $c_1(a,T)$ to one loop. The diagrammatic form of this condition is shown in\fig{tadpoles} and evaluates to,
\begin{align}\label{defc1}
c_1(a,T)&=\frac{T}{D(0)}\sum_n\int \frac{d\bk}{(2\pi)^2}[D_{\pm+}(0)\tr \bV_+^{(n)}(\tilde \bG_{++}^{|(n)}(\bk)+\bG_{++}^{|a|(n)}(\bk))+D_{\pm-}(0)\tr \bV_-^{(n)}(\tilde \bG_{--}^{|(n)}+\bG_{--}^{|a|(n)}(\bk))]\nonumber\\
&=c_1(\infty,T)-T\sum_n\int_0^\infty \frac{k dk}{2\pi}  \kappa\left( \frac{\bar r^2}{e^{2\kappa a}-\bar r^2}+\frac{\bar r^2}{e^{2\kappa a}-\bar r^2}\right)\nonumber\\
&=c_1(\infty,T)-\frac{\partial}{A\partial a} F_T^\parallel(a)\ ,
\end{align}
where $c_1(\infty,T)$ is the (infinite) one-interface contribution that does not depend on the separation $a$. The interpretation of \equ{defc1} is straightforward and could have been anticipated: for $\vev{h}\neq 0$, the separation $a$ is redefined at one loop. Since
\bel{changea}
F_T^\parallel(a)-\int_A d\bx h(\bx)\frac{\partial}{A\partial a} F_T^\parallel(a)\approx F_T^\parallel(a-\vev{h})\ .
\ee
To leading order in $\vev{h}$, the $c_1$-counterterm arises from the free energy of two parallel flat interfaces at separation $a_B$, where $a=a_B+\vev{h}$, is the separation at which \equ{vev0} holds. 

The $a$-independent but temperature-dependent contribution from $c_1(\infty,T)$ similarly  is the difference in free energy due to a shift of a flat interface by $-\vev{h}$. The bulk contribution to the free energy density thereby increases by,
\begin{align}\label{freeshift}
c_1(\infty,T)&=-\frac{T}{4}\sum_n\int_0^\infty \frac{k dk}{2\pi} \tr(V^{(n)}_+ \tilde \bG^{|(n)}_{++}(k)+V^{(n)}_- \tilde \bG^{|(n)}_{--}(k) )\nonumber\\
&=-\frac{T}{2}\sum_n(\eps(\zeta_n)-1)\int_0^\infty \frac{k dk}{2\pi}\left(\frac{\kappa_\eps\kappa-k^2}{\eps\kappa+\kappa_\eps}+\frac{\zeta^2}{\kappa_\eps+\kappa}\right)\nonumber\\
&=T\sum_n\int_0^\infty \frac{k dk}{2\pi} (\kappa-\kappa_\eps)=\frac{1}{V}(F_T^\gamma[1] -F_T^\gamma[\eps])
\end{align}
 where $F_T^\gamma[\eps]/V$ is the free energy density of a photon gas in a homogeneous medium  with permittivity $\eps(\zeta)$. The difference in free energy density in the dielectric and in vacuum depends on the permittivity $\eps(\zeta)$. For the plasma model with $\eps(\zeta)=1+(\oP/\zeta)^2$, this separation-independent contribution to the free energy is,
\bel{deltaFgamma}
(F_T^\gamma[1] -F_T^\gamma[\eps])\frac{A\vev{h}}{V}=A \vev{h}\big[c_1(\infty,0)-\frac{T^4\pi^2}{45}+\frac{T^2\oP^2}{\pi^2}\sum_{n=1}^\infty \frac{K_2(n\oP/T)}{n^2}\big]\ ,
\ee
where  the modified Bessel function $K_2(x)$ is normalized to $K_2(x\sim 0)\sim 2/x^2$.  The generally infinite constant $c_1(\infty,0)$ does not depend on temperature nor on the separation $a$. It is sensitive to the behavior of $\eps(\zeta)$ at energies $\zeta\gg \oP$. Estimating this contribution to the free energy in the framework of the low-energy effective theory is meaningless since the loop integral is dominated by momenta and energies  $k,\zeta\gg \oP$. For the sake of completeness,  this formal contribution with a proper time cutoff $\beta$ is,
\bel{cinfty}
c_1(a\sim\infty,T=0)=-\frac{1}{16\pi^2}\int_\beta^\infty \frac{d\lambda}{\lambda^3} (1-e^{-\lambda \oP^2})\ .
\ee
It is a quadratically and logarithmically UV-divergent constant contribution to the total energy of the system. It may be absorbed in the counter term $c_0$ and in the absence of gravitational interactions has no physical implications.  

\subsection{The Complete Low-Energy Effective Field Theory}
Since the Greens-function $\bG^\parallel$ of parallel interfaces  as well as the counter terms are invariant under transverse translations, the partition function $Z_T(\vec P=0, h)$ defined in  \equ{genfunction} for vanishing polarization sources is a functional of the roughness profile $h(\bx)$ with translation-invariant coefficients. We thus can use \equ{evalF} to evaluate it using the correlation functions of the profile $h(\bx)$ rather than the profile itself. We therefore have that,
\bel{Zeval}
 Z_T[\vec{ P}=0, h]=Z_T[\vec P=0, \frac{\delta}{\delta \alpha}] \left. Z_h[\alpha]\right|_{\alpha=0},
\ee
with $Z_\alpha[h]$ defined by\equ{Zh}. The complete generating functional of the Gaussian model we are considering thus is,
\bel{genfull}
 {\cal Z}_T[\vec{P},\alpha]:=\exp\Big[-\frac{1}{T}(F_T^\parallel(a)+\delta F[\delta/\delta\alpha])\Big]\prod_n \exp\Big[-\frac{1}{T}({\cal V}_n[\delta/\delta\alpha]+\delta{\cal V}^h_n)\Big]\exp\Big{[}\frac{T}{2}\{ \bP_n\vert \bG^{\parallel(n)}\vert  \bP_n\}+\frac{1}{2}(\alpha|D_2|\alpha) \Big{]}\ ,
\ee
with $(\alpha|D_2|\alpha)$ given by \equ{scalar2d}. The partition function of \equ{Zeval} is just ${\cal Z}_T[\vec{P}=0,\alpha=0]$. From the point of view of Euclidean field theory, \equ{genfull} promotes the roughness profile $h(\bx)$ to a field on the two-dimension (planar) subspace that is coupled to a vector field in $\mathbb{R}^3\times S_1$. Correlation functions of $h(\bx)$ are obtained by functional differentiation of\equ{genfull} with respect to the scalar source $\alpha$ and ${\cal Z}$ defines the loop-expansion in the usual manner. The main difference to ordinary field theory is that all correlation functions of $h(\bx)$  are prescribed and counter-functions enforce the absence of any corrections to them at $T=0$ and $a\sim\infty$. The low energy effective field theory encoded by \equ{genfull} evidently is not renormalizable -- new counter terms (functions) are required at each order of the loop expansion. The three counterterms $c_1$, $c_2$ of $\delta F[h]$ and $\delta \bV^h$ suffice at the 1-loop level since only the connected two-point functions and $\vev{h}$ are superficially UV-dominated if $D_2(0)=\sigma^2$ is finite.  

Instead of employing the Green's function approach, one can derive the loop corrections to the free energy from\equ{genfull}. The Casimir free energy to one loop is the same in both approaches. However, the generating functional  \equ{genfull} of the low-energy effective theory has  conceptual and methodical advantages: once the set of counter-terms is determined, the field theory yields consistent low-energy results not just for the Casimir energy, but  for the scattering matrix as well. No ad-hoc arguments and procedures are required to cancel uncontrolled high-energy loop corrections and the necessity of the counter terms and their interpretation is readily apparent.    

\section{Numerical Investigations}
\label{secNum}
We numerically investigated the correction  $\Delta F_T^\text{Cas}(a)$ to the Casimir free energy given in \equ{rough2F} due to the roughness of an interface. To order $\sigma^2$ this correction is linear in the roughness correlation function and one may define\cite{Neto2005} a response function $R_T(q, a)$,
\bel{defR}
\Delta F_T^\text{Cas}(a)=\int_0^\infty \frac{q dq}{2\pi}R_T(q, a) D(q)\ ,
\ee
 that does not depend on $D(q)$. Analytical expressions for $R_T(q, a)$ are obtained by changing the integration variable from $\bk'$ to $\bq=\bk'-\bk$  in Eqs.~(\ref{seagull}),~(\ref{fig2b1}),~(\ref{fig2c})~and~(\ref{fig2d}). The corresponding expressions are given in App.~\ref{appD}. For clarity and  to compare with earlier investigations, we in the following present numerical results for $T=0$ only. Temperature corrections are sizable only when $2\pi aT\gtrsim 1$. For gold surfaces at $300^o$K,  temperature corrections become important at separations of the order of microns - a distance at which perturbative roughness is irrelevant. 
\begin{figure}
\includegraphics[scale=0.8]{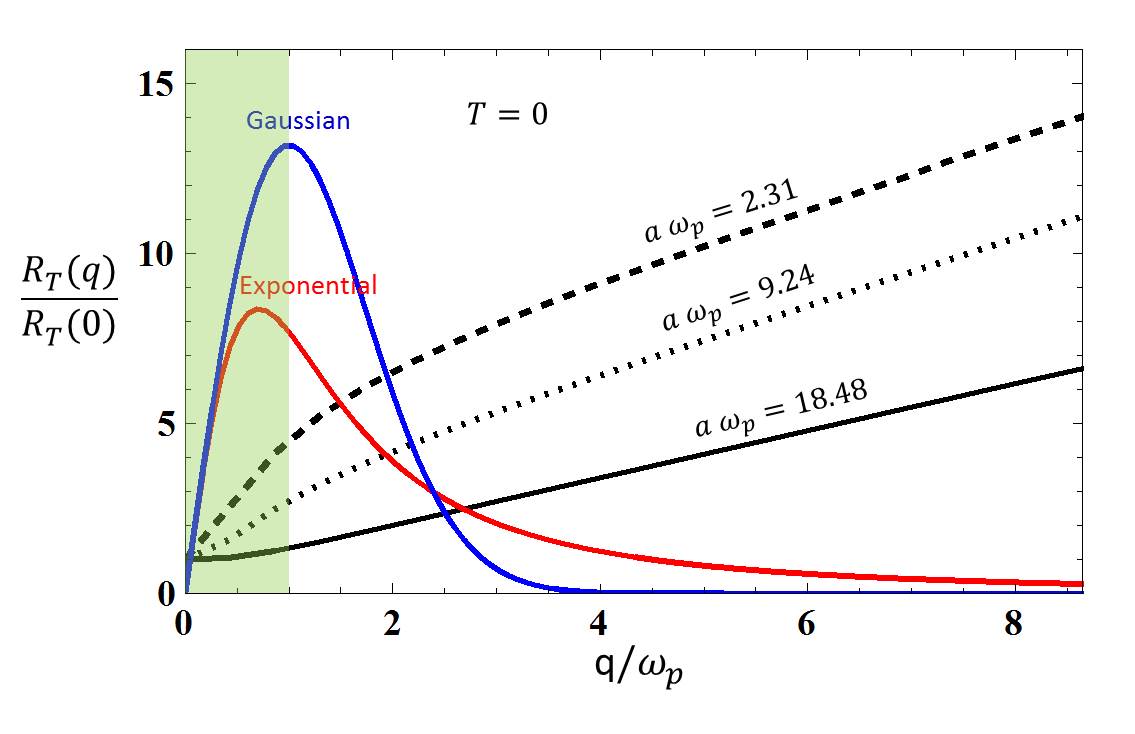}
\caption{The dimensionless normalized response $\rho(q,a) =R_T(q,a)/R_T(0,a)$ without counter potential $\delta\bV^h=0$ for the permittivity $\eps(\zeta)=1+(\oP/\zeta)^2$ to leading order in $\sigma^2$ at $T=0$. The dependence on $q/\oP$ of this ratio of the roughness response function $R_T(q,a)$ (defined by \equ{defR}) is shown for $a\oP=2.31(-~-),9.24(\cdot\cdot\cdot\cdot\cdot)\ \text{and}\ 18.48(-\!\!\!-\!\!\!-\!\!\!-)$. For the plasma frequency $\oP=\oP(\text {Au})\sim 0.046 \text{nm}^{-1}$, this normalized response without counter potential is identical with that obtained by Ref.~\cite{Neto2005}. [For $\oP=0.046\text{nm}^{-1}$ the curves here corresponds to those of Fig.~4 in Ref.~\cite{Neto2005} at separations  $a=50, 100,$~and~$200 \text{nm}$.]  Note the change in behavior and subsequent linear rise in the region $q \oP\gtrsim1$. The region $q \oP\lesssim1$ where the effective low-energy theory is valid is shaded light green. We superimpose typical integration densities for the response function in\equ{defR}: the momentum space function $q D(q)$ for Gaussian and exponential 2-point roughness correlation with $l_c=1/\oP$. The roughness correction to the Casimir energy with exponential correlation diverges logarithmically and even for Gaussian roughness correlation the (unshaded)  region $q/\oP>1$ contributes significantly in this uncorrected case. Note that for a gold surface the correlation length here is $l_c=1/\oP(\text{Au})\sim 21\text{nm}$.}  
\label{rho}
\end{figure}
\subsection{The Response with and without Counter Term}
 \fig{rho} gives the normalized response when the counterterm of \equ{fig2c} is omitted as a function of the dimensionless variable $q/\oP$. The low-energy theory is in the shaded  momentum region $q/\oP<1$.  Note the linear rise of the low-energy  response function for all separations $a$ in the uncontrolled region $q/\oP\gg 1$. The integration weight $qD(q)$ for Gaussian and exponential roughness correlation with a typical correlation length $l_c\sim 1/\oP$ is superimposed.  A sizable contribution to the roughness correction in\equ{defR} evidently is due to loop momenta $q>\oP$ for which low-energy expressions are unreliable.  

\begin{figure}
\includegraphics[scale=0.8]{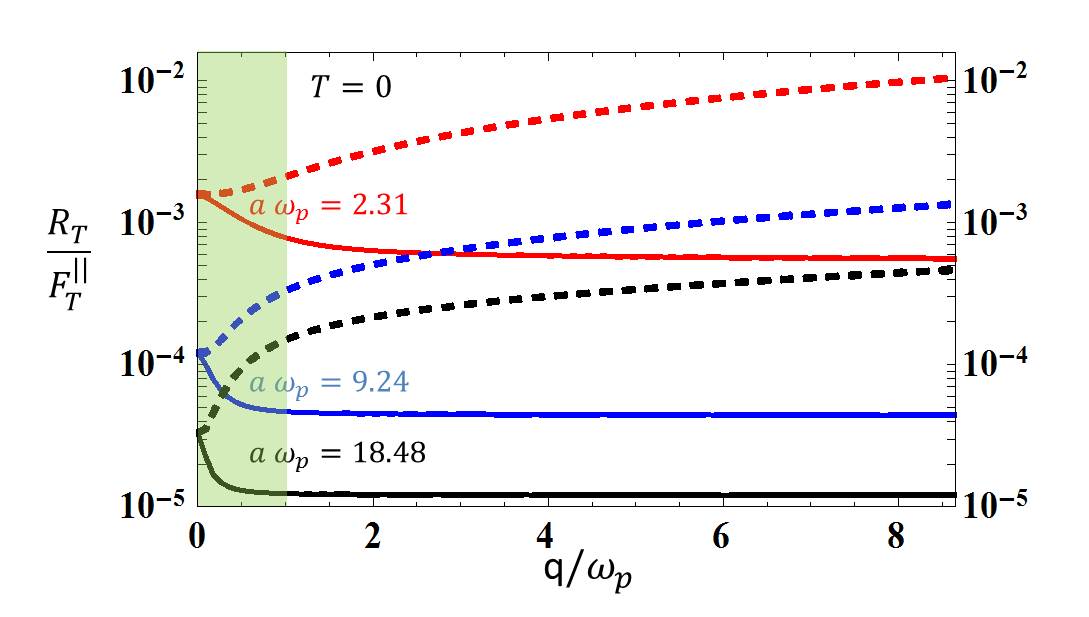}
\caption{(Color online) The ratio  $ R_T(q,a)/F_T^\parallel(a)$ of the roughness response function to the Casimir energy of flat parallel plates  at $T=0$ with (solid) and without (dashed) counter potential $\delta\bV^h$ with $g^2=1$. The permittivity $\eps(\zeta)=1+(\oP/\zeta)^2$ is characterized by the plasma frequency $\oP$. The dependence on $q/\oP$ of the ratio is shown for $a \oP=2.31\text{(top,red)},9.24\text{(middle,blue)}\ \text{and}\  18.48\text{(bottom,black)}$. For $\oP=0.046 \text{nm}^{-1}\sim \oP(\text {Au})$ the normalized response without counter potential (dashed) is identical with that of Fig.~3 in Ref.~\cite{Neto2005} at separations  of $a=50, 100,$~and~$200 \text{nm}$.  Note that the renormalized roughness response is monotonically decreasing and approaches a constant at large momenta that is a factor of 2-3 smaller than the response at $q=0$. Most of the correction to the Casimir energy in this case arises from the shaded integration  region $q/\oP<1$ where the low-energy description is valid.}  
\label{response}
\end{figure}

Inclusion of the counter potential gives a constant high-momentum response. \fig{response} shows the response functions with and without the counterterm contribution  of\equ{fig2c}. With the same model for the bulk permittivity of gold, the response function shown in Fig.~3 of  Ref.~\cite{Neto2005} is reproduced when the counter-potential is omitted.  Inclusion of the counter potential gives a constant high-momentum response and the correction to the Casimir (free) energy is of order $\sigma^2$.  Note that with $g^2=1$ the response at $q=0$ does not change. 

\begin{figure}
\includegraphics[scale=0.7]{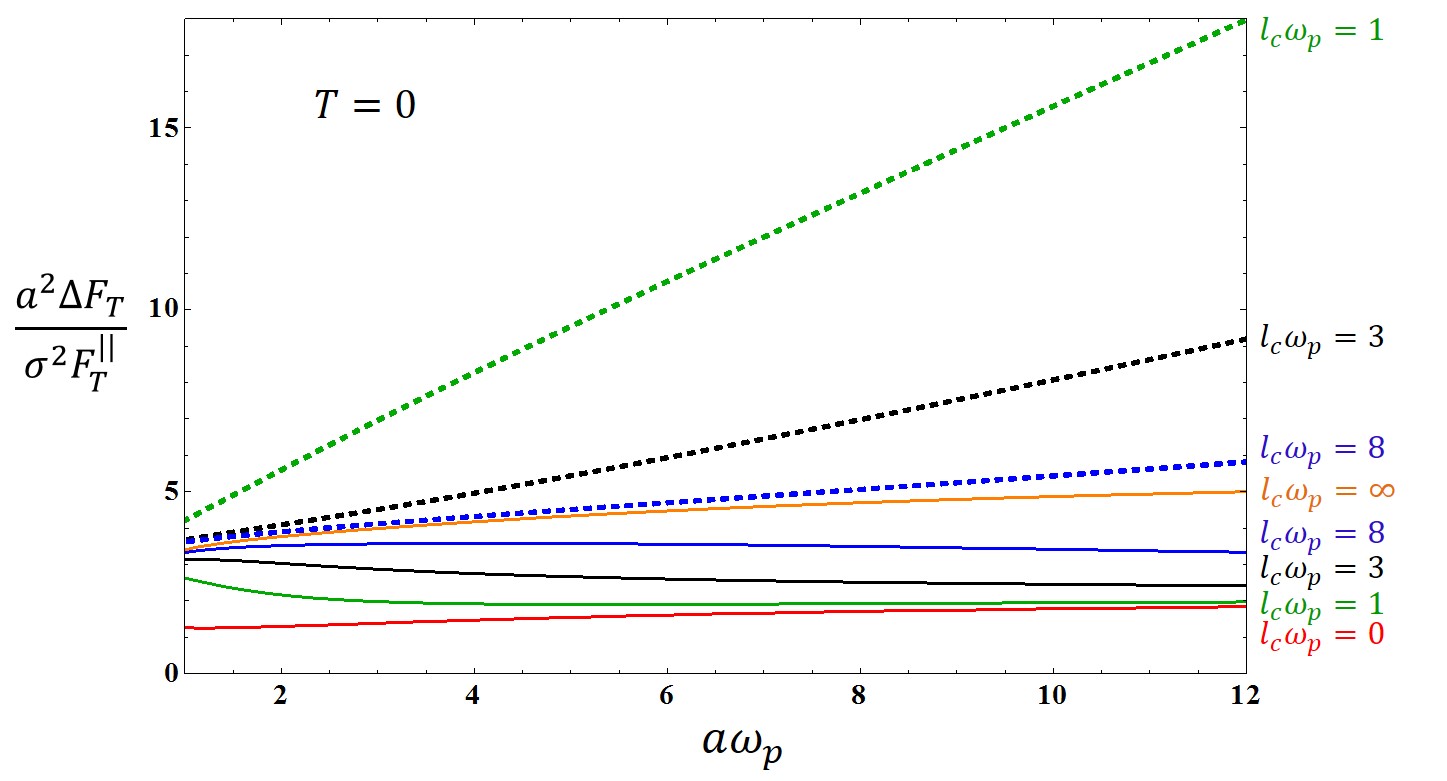}
\caption{(Color online) The dimensionless ratio  $(a^2/\sigma^2) \Delta F^\text{Cas}_T(a)/ F_T^\parallel(a)$ of the roughness correction to the Casimir energy of two parallel flat interfaces at $T=0$. The calculation is to leading order in $\sigma^2/a^2$ for a plasma-model permittivity with plasma frequency $\oP$  for Gaussian roughness with correlation length $l_c$.  Dashed curves give the ratio as a function of $a\oP$ without the counter term contribution of \equ{fig2c} whereas solid curves give the ratio when this counter term with $g^2=1$ is included. Curves of the same color correspond to the same value of $l_c\oP$. From the top: $l_c\oP=$ 1 (green, dashed), 3 (black,dashed), 8 (blue,dashed), $\infty$ (orange), 8 (blue,solid), 3 (black,solid), 1(green,solid) and 0 (red,solid). Note that the  $l_c\rightarrow 0$ curve (red) is a lower bound that exists only in the renormalized case. The counter term vanishes in the PFA limit $l_c\rightarrow\infty$ (orange), and this limit is the same for both. Whereas the PFA is an upper bound for the magnitude of the roughness correction when the counter potential is included, it is a lower bound without. The ratio of the roughness correction to the PFA at finite $l_c$   approaches a finite value at large separations when the counter term is included whereas it otherwise increases indefinitely. The roughness correction in the subtracted case at large separations is less than $50\%$ of the PFA-prediction when $l_c\lesssim 1/\oP$.  Except for $l_c=0$, the roughness correction approaches the PFA estimate at sufficiently small separation, but it quickly decreases and approaches the lower bound for $l_c\oP<1$. }  
\label{Roughcorr}
\end{figure}

The correction  to the Casimir energy at $T=0$  for Gaussian roughness with and without inclusion of the counter term of \equ{fig2c} is shown in\fig{Roughcorr}. Whereas the PFA-limit $l_c\rightarrow \infty$ coincides for both cases, the behavior is remarkably different at finite $l_c$. Including the counter term of \equ{fig2c} the roughness correction to the Casimir energy \emph{decreases} in magnitude  for decreasing correlation length and approaches a finite (uncorrelated) limit  for $l_c\rightarrow 0$.
Roughness increases the Casimir force but the PFA is an upper bound in this case. The ratio of the roughness correction to the PFA furthermore approaches a constant, $l_c$-dependent,  value with increasing separation rather than increasing indefinitely  as in the unsubtracted case (for exponential roughness, the roughness correction without the counter term of \equ{fig2c} would diverge at any separation and for all $l_c$). Let us also note that for $l_c\lesssim 1/\oP$ the roughness correction at large separations is less than 50\% of the PFA prediction. Although we here are considering only perturbative roughness corrections,  the suppression at large separations for $l_c\lesssim 1/\oP$ is of a similar magnitude as that observed 
\cite{Decca2013} for machined profiles with correlation length $l_c\sim 1/\oP$.  

\subsection{(In)sensitivity on High Momentum Components of  the Roughness Correlation}
\begin{figure}
\includegraphics[scale=0.8]{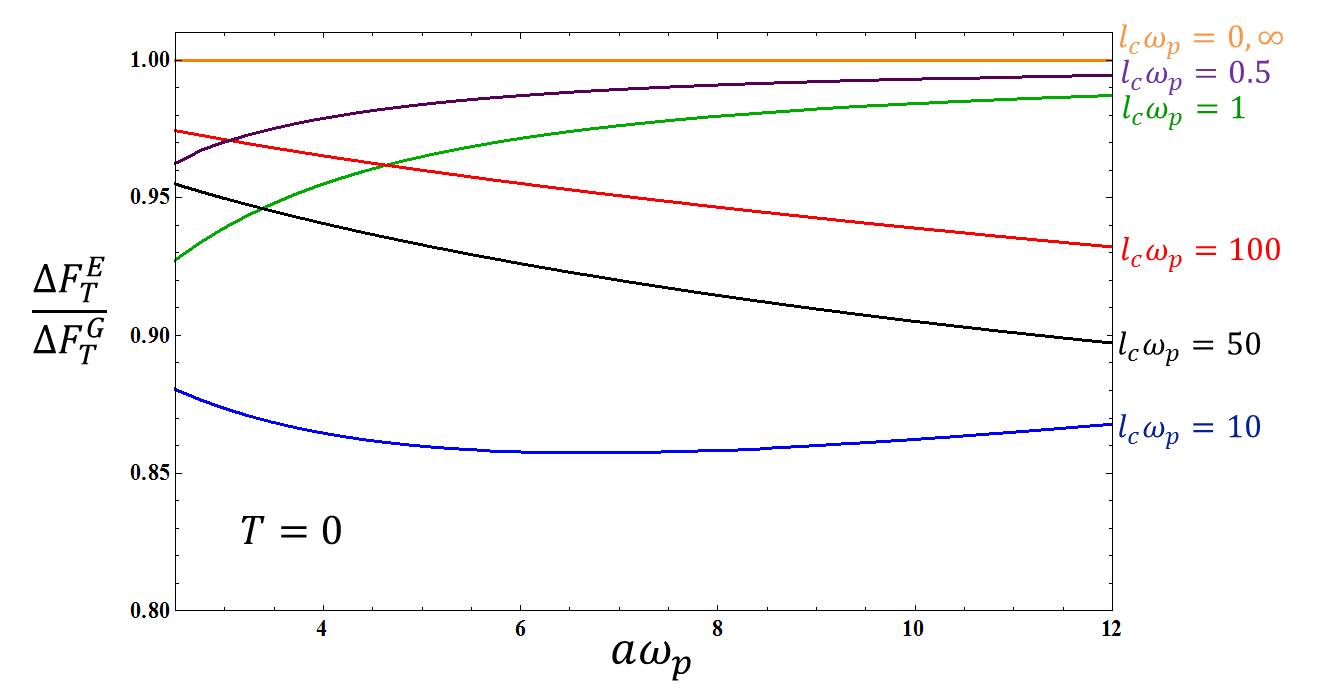}
\caption{(Color online) The dimensionless ratio  $\Delta F^E_T(a)/ \Delta F^G_T(a)$ of the roughness correction to the Casimir energy for exponential(E) and Gaussian(G) roughness with the same correlation length $l_c\oP$ as a function of the dimensionless separation $a\oP$. $g^2=1$ and a plasma-model permittivity characterized by the single plasma frequency $\oP$ was assumed.  The roughness correlation functions are those of \equ{ExpD}(E) and \equ{GaussD}(G). In the PFA ($l_c\rightarrow\infty$) and uncorrelated ($l_c\rightarrow 0$) limits the corrections coincide but differ by up to $15\%$ at some separations. For the same variance $\sigma^2$ and correlation length $l_c$, the roughness correction with exponential correlation is always smaller than with Gaussian correlation. Note that the two types of roughness correlation approach the PFA quite differently: at large separations the corrections still differ by over $5\%$ even for $l_c\oP\sim 100$.}  
\label{ratGE}
\end{figure}

The counter potential $\delta \bV^h$ was introduced to correct for uncontrolled high-momentum contributions to loop integrals with the help of  phenomenological input. We therefore investigated the sensitivity of the roughness correction to the correlation function $D(q)$ numerically. \fig{ratGE} shows the ratio of the correction for Gaussian- and for exponential- roughness of the same correlation length $l_c$. The two are identical for $l_c=0$ and $l_c\sim\infty$ (PFA) at any $a\oP$. The (dimensionless) ratio of these corrections never drops below $85\%$ for any separation $a\oP$ and correlation length $l_c\oP$.  Without counter potential this ratio is infinite. Exponential roughness always gives a smaller correction than Gaussian roughness of the same correlation length and variance. The two correlation functions provide rather similar descriptions of low energy scattering and the low-energy effective theory with counter potential depends only weakly on their (very different) behavior at high momenta. 

\subsection{Comparison with Experiment}
\label{ExpComp}
The low energy theory for electromagnetic interactions with rough surfaces ultimately must be compared to experiment. Unfortunately only very few studies are dedicated to the systematic investigation of Casimir forces between rough surfaces. Many employ non-isotropic machined surfaces with rather large $\sigma/a$-ratios\cite{Chan2008, Decca2013} that are not accessible perturbatively.  Nevertheless, these experiments qualitatively contradict the predictions of exact calculations, that essentially any kind of roughness tends to increase the Casimir force above the PFA estimate. 
A notable exception is a series of investigations of isotropically rough surfaces by Palasantzas et al.\cite{Zwol20071,*Zwol20081,Pala2008}. For sufficiently rough surfaces, this group does observe (see Fig.~3 of Ref.~\cite{Pala2008}) an increase of the Casimir force by 200-400\% at small separations. This sharp increase in the force was attributed to particularly high islands of the surface profile that can also be seen in some of the AFM scans of the gold surfaces. The pronounced effect of such islands is beyond the scope of a perturbative analysis and was explained  by a semi-empirical approach\cite{Broer2012} based on the PFA.  

\begin{figure}
\includegraphics[scale=0.8]{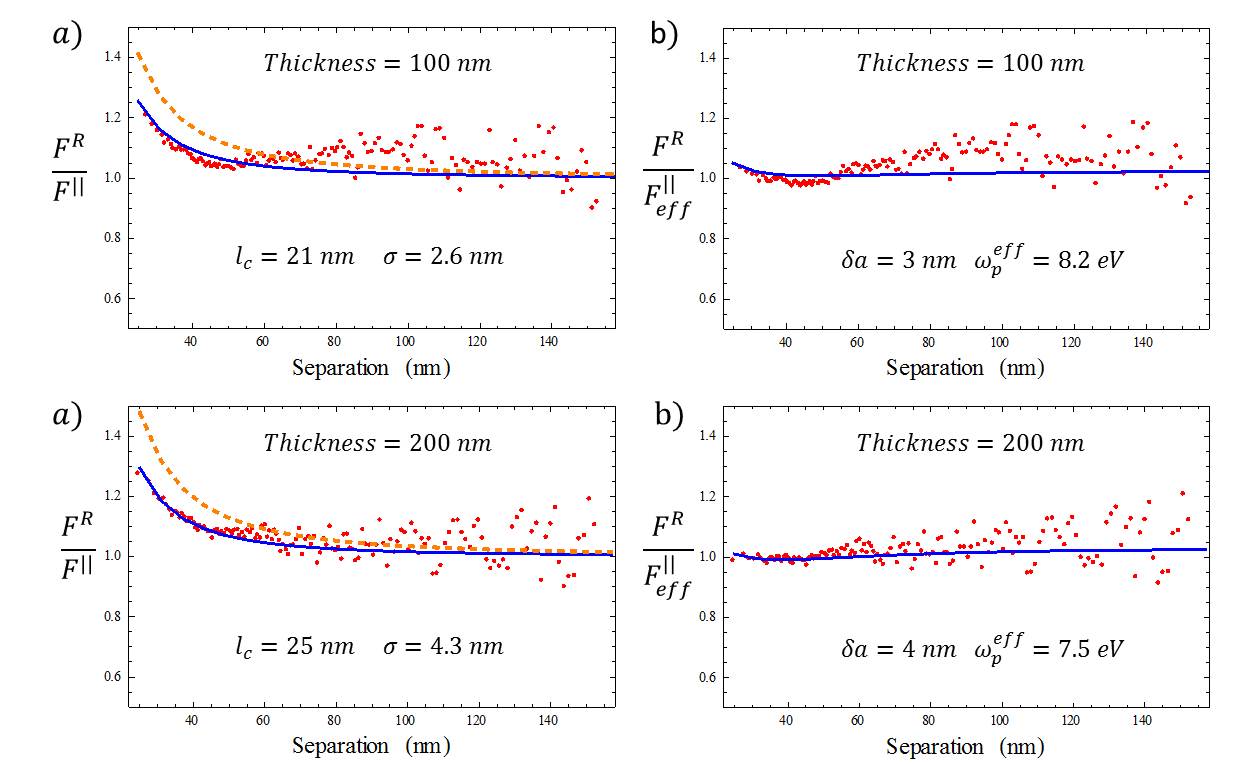}
\caption{(Color online) The dimensionless ratio  $\rho(a)$ defined in \equ{ratioDer} of the Casimir force between a rough gold-coated sphere and a rough gold-coated plate to the Casimir energy between ideal dielectric flat plates. The experimental data is from Ref.~\cite{Pala2008}. The thickness of the gold coating on the flat plate is $100$nm (upper graphs) and $200$nm (lower graphs). An exponential roughness correlation and a Drude parametrization of the permittivity is assumed. The standard deviation and correlation length for the sphere's profile is  $\sigma^\text{Sph}\sim 8$nm  and $l^{Sph}_c\sim 33$nm.  a) The ratio of the force on the rough plate to the Casimir force between a gold-coated flat plate and a smooth sphere at the same mean separation.  A Drude parametrization of the permittivity with $\oP=9$eV, $\gamma=0.045$eV was used. (Red) dots is the ratio for experimental data of Ref.~\cite{Pala2008}. The  measured force on the $100$nm thick plate was  multiplied by a correction factor of $0.94$ (see text for details).  The solid (blue) line is our best theoretical fit to this ratio with the indicated parameters for the roughness correlation function of the plate in \equ{combD}. Note that the $\sim 30\%$ enhancement at separations $a\sim 20$nm is well reproduced for both films. The dashed line gives the PFA for roughness of the same total variance. b) The ratio of the force on the rough plate to that between a smooth sphere and a flat plate at the separation $a-\delta a $.  The  indicated $\oP_\text{eff}$ for the effective permittivity of the flat plate was obtained  from  ellipsometric measurements\cite{Pala2008} on the rough ones. We assumed the same effective plasma frequency  $\oP^\text{Sph}_\text{eff}=7.5$eV for the sphere as for the (similarly rough) $200$nm film.  The solid (blue) line gives the ratio to the force on the effective flat plate and sphere for the same force including the roughness corrections shown in a).  Note that this ratio of the force with roughness corrections to that between a flat plate and smooth sphere with the measured reflection coefficients at a reduced separation is close to unity for all separations.}  
\label{CompPala}
\end{figure}

However, gold films with $100$nm and $200$nm thickness of relatively low roughness appear to be almost free of such buildup effects. At small separations the force in these cases is smaller than the PFA prediction. In \fig{CompPala} we compare the low-energy theory to the measurements of Ref.~\cite{Pala2008} on these thin films.  The experiments measure the force between a gold-coated sphere and a gold-coated plate. Both surfaces are rough, but their profiles are uncorrelated. For two parallel rough gold-coated plates the correction to the Casimir energy to leading order in $\sigma/a$ is that for a single rough plate with a roughness correlation that is the sum of the roughness correlations functions of the sphere and the flat plate,
\bel{combD}
D(q)=D^\text{plate}(q)+D^\text{sphere}(q)\ .
\ee
We use Derjaguin's PFA approximation\cite{Derjaguin19581} to correct for the curvature of the sphere of radius $R=100\text{$\mu$m}\gg a$. The force $f_T(a)$ at temperature $T$ between the sphere and a plate with (closest) separation $a$ in this approximation is,
\bel{Der}
f_T(a)= 2\pi R F_T^\text{Cas}[a]/A\ ,
\ee
where $F_T^\text{Cas}[a]/A$ is the Casimir free energy per unit area (not the pressure) of two parallel rough plates. Due to the large radius of the sphere, this is an excellent approximation for separations $a<200\text{nm}\sim R/500$. 
\fig{CompPala}a gives the ratio $\rho(a)$ of this force to the Casimir energy per unit area $F_T^\parallel[a]/A$ of two \emph{flat} parallel gold plates with separation $a$,
\bel{ratioDer}
\rho(a):=\frac{f_T(a) A}{2 \pi R F_T^\parallel[a]}
 =\frac{F_T^\text{Cas}[a]}{F_T^\parallel[a]}=1+\frac{\Delta F_T^\text{Cas}[a]}{F_T^\parallel[a]}\ ,
\ee
at $T=0$. The experimental Casimir force for the rough sphere and plate at separations $\sigma\ll a< l_c$  is up to 30\% greater than the Casimir energy for flat plates.   

Since we do not differentiate between contributions from high and low peaks of the roughness profile and only use a single correlation function, all standard deviations of Ref.~\cite{Pala2008} were multiplied by a factor of $1.7$. We used $\sigma^\text{Sph}=8$nm, $\sigma^\text{100}=2.6$nm and $\sigma^\text{200}=4.3$nm for the coatings of the sphere, $100$nm and $200$nm thick films respectively.  These standard deviations also approximately correspond to those estimated from the AFM-scans of these surfaces (see Fig.~1 in Ref.~\cite{Pala2008}). The correlation lengths $l_c^\text{Sph}=33$nm, $l_c^\text{100}=21$nm and $l_c^\text{200}=25$nm are those of Ref.~\cite{Pala2008}.  The ratio $\rho(a)$ for the $200$nm thick film is well reproduced by the low-energy theory with exponential as well as with Gaussian correlations. We only show the result for exponential roughness in \fig{CompPala}, but the fit for Gaussian roughness is of similar quality. For comparison we show the roughness correction in PFA for the same standard deviations.

The ratio $\rho(a)$ is close to unity at larger separations $100\text{nm}<a<150\text{nm}$ where roughness corrections are relatively small. While this on average is approximately observed for the  $200$nm film, the ratio for the $100$nm film is systematically about $6\%$ above unity at  larger distances. To correct for this (unexplained) discrepancy we multiplied the force observed on the $100$nm thick film by $0.94$ before\footnote{While this correction factor is \emph{ad hoc}, we would like to point out that the ratios of \fig{CompPala} are less forgiving than logarithmic depictions of the data. The experimental error probably increases sharply at larger separations simply because the force is rapidly decreasing in magnitude.} comparing with theory.  

From a practical point of view the comparison in \fig{CompPala}b with the Casimir energy of two parallel flat plates at a slightly smaller separation $a_\text{eff}=a-\delta a$ perhaps is more useful. The Drude-model permittivity describing reflection off these effective flat plates in Ref.~\cite{Pala2008} was obtained from ellipsometric measurements on the rough surfaces. We merely adjusted $\delta a$ for the best fit. \fig{CompPala}b shows that  effective flat surfaces at a reduced separation $a-\delta a$ reproduce the low-roughness data remarkably well. [The force data of the $100$nm film was multiplied by the same correction factor of $0.94$ as in the graph of \fig{CompPala}a. ] Since ellipsometric measurements on thin films are quite standard, this observation essentially reduces low-roughness corrections to Casimir energies to a determination of the optimal shift $\delta a$. Instead of measuring the absolute average distance between the profiles of two rough surfaces (in itself a delicate procedure that involves a number of corrections), we suggest that precision Casimir studies with low-roughness surfaces simply determine an\emph{effective} separation for flat plates with the measured  (perpendicular) reflection coefficients. \fig{CompPala}b is evidence that the data at small separations robustly determines this distance to better than $1$nm, at the same time all but eliminating the need for roughness corrections. 

\section{Conclusion}
\label{Concl}
We obtained roughness corrections to low-energy scattering and the Casimir free energy in the framework of Schwinger's effective theory of low energy electrodynamics. The energy scale in this theory is the plasma frequency $\oP\sim0.046 \text{nm}^{-1}\sim 9\text{eV}$ of typical materials like gold. We found that roughness corrections generally include large contributions from high momentum excitations. Evaluating them in the low-energy framework is inconsistent and  notoriously unreliable.  We emphasize that this is not a limitation of the perturbative approach developed here: exact (numerical) solutions of a model can also only be as accurate as the model itself.  The Casimir energy of short-wavelength periodic rectangular profiles for instance involves momenta at which a description in terms of the bulk permittivity of the material breaks down and the mathematically exact analysis of such a model can lead to physically erroneous conclusions. Using the bulk permittivity to describe scattering off profile structures with sizes of the order of the inverse plasma frequency or smaller (about $25$nm for gold) is not justified. Effects due to roughness on the scale of the plasma frequency generally are grossly overestimated  by the uncorrected low-energy theory. This has been experimentally verified for machined profiles with a period $\lambda\lesssim 2 \pi/\oP$: the exact calculations\cite{Lambrecht2008,Davids2012} for such profiles tend to over-estimate the observed\cite{Decca2013} Casimir force by factors of 2-3. 

We presented a perturbative analysis of roughness corrections based on a low-energy effective field theory that employs counter-terms to correct for uncontrolled high-momentum contributions. The counter terms subtracts high-momentum contributions to loop integrals at the cost of phenomenological input. Apart from correlations of the roughness profile itself, we in addition modeled the averaged single-interface scattering matrix at vanishing transverse momentum by the plasmon contribution. To leading order in the roughness variance $\sigma^2$ this semi-empirical ansatz depends on a single coupling constant $g^2$.  Consistency of the low-energy theory and the existence of an ideal metal limit at any correlation length constrains this dimensionless coupling to $g^2=1$ at low energies (see \equ{g1}).  The resulting low-energy theory is free of high-momentum contributions to one-loop integrals, approaches the PFA for $l_c\sim\infty$ and has a finite ideal metal limit for any $l_c$. It is relatively insensitive to the high-momentum behavior of the roughness correlation function and has a drastically different but more transparent dependence on $l_c$ than the uncorrected model.  Instead of large (infinite) differences, roughness correlation functions that differ only at high momenta now give similar low-energy predictions. Roughness of shorter correlation length no longer increases the Casimir force (indefinitely). Instead the magnitude of the force decreases with decreasing correlation length and approaches a finite lower bound for uncorrelated roughness.   

Although the coupling $g^2$ in the plasmon contribution to the counter-term potential \equ{dtV}  was constrained to $g^2=1$ by selfconsistency and the existence of certain limits of the effective low-energy theory,  this is a \emph{model} for the roughness contribution to the average scattering matrix at low transverse momenta. It may be phenomenologically preferable to parameterize empirical data for this component of the scattering matrix instead. However, there is some evidence that the plasmon describes low-energy scattering due to roughness reasonably well. It in this sense is a reasonable model for the leading roughness correction that is relatively simple and consistent with the low energy theory.       

Interestingly the PFA is accurate at small separations only for $l_c\gtrsim1/\oP$ and at large separations may overestimate the correction to the force by up to 250\% (see \fig{Roughcorr}).  For $l_c\lesssim 1/\oP $ the roughness correction to the Casimir energy is significantly (a factor $\sim 1/2-1/3$) below the PFA prediction at all but the smallest separations. The ratio remains approximately constant for $a\sim\infty$ and does not increase with increasing separation as in the uncorrected model.   Although we considered only isotropic roughness profiles,  it perhaps is interesting that the reduction of the correction compared to the PFA prediction by a factor of $2$ for $l_c\sim1/\oP$ is of the same order of magnitude as the experimental reduction in the overall force observed\cite{Decca2013} by experiments with corrugated rectangular wave profiles.      

The Casimir energy of low-roughness profiles was found to be essentially that of flat plates with the \emph{measured} reflection coefficients at a distance that is  slightly smaller than the mean separation of the interfaces. The change in separation is less than the standard deviation of the rough profile. Although the precise value of this shift depends on properties of the profile, this observation enables one to empirically correct for (low-level) roughness  and accurately calibrate the effective separation in the plate-sphere geometry. 

For conceptual reasons we here derived all expressions for the Casimir free energy at finite temperature, but only investigated implications of this theory at $T=0$. We intend to extend the numerical investigations to finite temperature in the future. Although the roughness correction at finite temperature is not expected to change at small separations, the regime $1<a/l_c<a T$  where temperature and roughness corrections are of similar importance could be of some interest.At this point we only wish to observe that the summands in all expressions at finite temperature are finite when $\zeta\rightarrow 0$ for any reasonable permittivity function (Drude- or plasma-model).  Predictions of this low-energy effective field theory at temperatures $2\pi T>\oP\sim 2\times 10^4\  {^o}K$ nevertheless would be meaningless.

\acknowledgments
We would like to thank G.~Palasantzas for giving us access to the experimental data of his group. Discussions with K.V.~Shajesh and  Junming~Liu provided   insights that are gratefully acknowledged. H.-Y. W. enjoyed the support and hospitality of the Lorentz Center in Leiden, Netherlands and the invitation to PASI2012 where some preliminary results were presented. This work was supported by NSF Grant  PHY-09-02054.

\appendix
\section{The Green's Dyadic for Three Flat Dielectric Slabs}
\label{appA}
In Schwinger's formalism\cite{Schwingerin1978} the parallel-plate Green's dyadic is determined by reduced  electric and magnetic Green's functions. In the coordinate system in which $\bk=(k,0)$ points along the +x axis, this Green's dyadic is,
\bal{gamma}
\bG^\parallel(k,z,z';\zeta, a)=\begin{bmatrix}
-\frac{1}{\eps_z}\frac{\partial}{\partial z}\frac{1}{\eps_{z'}}\frac{\partial}{\partial z'}g_{H} & 0 & -\frac{ik}{\eps_z\eps_{z'}}\frac{\partial}{\partial z}g_{H}\\
0 & \zeta^2 g_{E} & 0\\
\frac{ik}{\eps_z\eps_{z'}}\frac{\partial}{\partial z'}g_{H} & 0 &\frac{1}{\eps_z}\delta (z-z')-\frac{k^2}{\eps _z\eps_{z'}}g_{H}\\
\end{bmatrix}
\ea
where the $g_{E}$ and $g_{H}$ solve the differential equations,
\bal{rg}
\Big[-\frac{\partial^2}{\partial z^2}+k^2+\zeta^2\eps_z\Big]g_{E}(k ,z,z';\zeta)=\delta(z-z')\\
\Big[ -\frac{\partial}{\partial z}\frac{1}{\eps_z}\frac{\partial}{\partial z}+\frac{k^2}{\eps_z}+\zeta^2\Big]g_{H}(k ,z,z';\zeta)=\delta(z-z')\nonumber
\ea
One recovers the Green's function for arbitrary transverse momentum $\bk$  by   rotation about the $z$-axis,
\bal{rotation}
\bG^\parallel(\bk,z,z';\zeta, a)&=&\mathbf{R}\cdot \bG^\parallel(k=|\bk|,z,z';\zeta, a)\cdot \mathbf{R}^T\\
\mathbf{R}&=&\frac{1}{k} \left(
\begin{array}{ccc}
k_x & -k_y & 0 \\
k_y & k_x & 0 \\
0 & 0 & k
\end{array} \right)\nonumber
\ea
The solution to \equ{rg} in different regions of $z$ and $z'$ will be denoted,
\bal{geinfd}
g_{i}(k,z,z';\zeta)=\begin{bmatrix}
g_{i}^{ ++}(k,z>0,z'>0;\zeta) &g_{i}^{+-}(k,z>0,z'<0;\zeta) \\
g_{i}^{-+}(k,z<0,z'>0;\zeta) &g_{i}^{--}(k,z<0,z'<0;\zeta)\\
\end{bmatrix}\text{   with  }i=E\text{ or }H\ .
\ea
We divide the reduced Green's functions into $g_i^|$ for a single flat plate and its correction $g_i^{|a|}$ due to the presence of a parallel flat plate at a distance $a$: 
\bal{g}
g_{i}(k,z,z';\zeta, a)=g_{i}^|(k,z,z';\zeta) + g_{i}^{|a|}(k,z,z';\zeta, a)
\ea
\bal{gehinf}
&&g_{E}^|(k,z,z';\zeta)=\begin{bmatrix}
\frac{1}{2\kappa_2}(e^{-\kappa_{2}\vert z-z' \vert}-r_{2}e^{-\kappa_{2}(z+z')}) & \frac{1}{\kappa_2+\kappa_3}e^{\kappa_{3} z'-\kappa_{2} z} \\
 \frac{1}{\kappa_2+\kappa_3}e^{\kappa_{3} z-\kappa_{2} z'} & \frac{1}{2\kappa_3}(e^{-\kappa_{3}\vert z-z' \vert}+r_{2}e^{\kappa_{3}(z+z')})\\
\end{bmatrix}\\
&&g_{H}^|(k,z,z';\zeta)=\begin{bmatrix}
\frac{1}{2\bar\kappa_2}(e^{-\kappa_{2}\vert z-z' \vert}-\bar r_{2}e^{-\kappa_{2}(z+z')}) & \frac{1}{\bar\kappa_2+\bar\kappa_3} e^{\kappa_{3} z'-\kappa_{2} z} \\
\frac{1}{\bar\kappa_2+\bar\kappa_3} e^{\kappa_{3} z-\kappa_{2} z'} & \frac{1}{2\bar\kappa_3}(e^{-\kappa_{3}\vert z-z' \vert}+\bar r_{2}e^{\kappa_{3}(z+z')})\\
\end{bmatrix}\nonumber
\ea
\bal{geha}
&&g_{E}^{|a|}(k,z,z';\zeta, a)=\frac{r_{1}}{e^{2a\kappa_{3}}-r_{1}r_{2}}\begin{bmatrix}
\frac{1}{2\kappa_2}(1-r_{2}^2)e^{-\kappa_{2}(z+z')} & \frac{1}{\kappa_2+\kappa_3}( e^{-\kappa_{2} z-\kappa_{3} z' }+r_{2}e^{-\kappa_{2}z+\kappa_{3} z'})\\
  \frac{1}{\kappa_2+\kappa_3}( e^{-\kappa_{2} z'-\kappa_{3} z }+r_{2}e^{-\kappa_{2}z'+\kappa_{3} z} )&\frac{1}{2\kappa_3} (e^{-\kappa_{3} z}+r_{2}e^{\kappa_{3} z}) (e^{-\kappa_{3} z'}+r_{2}e^{\kappa_{3} z'})\\
\end{bmatrix}\nonumber\\
&&g_{H}^{|a|}(k,z,z';\zeta, a)=\frac{\bar r_{1}}{e^{2a\kappa_{3}}-\bar r_{1}\bar r_{2}}\begin{bmatrix}
\frac{1}{2\bar\kappa_2}(1-\bar r_{2}^2)e^{-\kappa_{2}(z+z')} &\frac{1}{\bar\kappa_2+\bar\kappa_3}( e^{-\kappa_{2} z-\kappa_{3} z' }+\bar r_{2}e^{-\kappa_{2}z+\kappa_{3} z'})\\
\frac{1}{\bar\kappa_2+\bar\kappa_3}(  e^{-\kappa_{2} z'-\kappa_{3} z }+\bar r_{2}e^{-\kappa_{2}z'+\kappa_{3} z}) & \frac{1}{2\bar\kappa_3} (e^{-\kappa_{3} z}+\bar r_{2}e^{\kappa_{3} z}) (e^{-\kappa_{3} z'}+\bar r_{2}e^{\kappa_{3} z'})\\
\end{bmatrix}\nonumber
\ea
Note that continuity of $E_x$, $E_y$, and $\eps E_z$ across the flat interface implies that of $g_E$, $g_H$, and $\frac{1}{\eps_z}\frac{\partial}{\partial z}\frac{1}{\eps_{z'}}\frac{\partial}{\partial z'}g_{H}$ are continuous as well. The components of  \equ{gamma} in different regions domains of $z$ and $z'$ are: 
\begin{align}\label{Ginf}
\tilde G_{xx}^{|}(k,z,z';\zeta)&=&-\frac{1}{\eps_z}\frac{\partial}{\partial z}\frac{1}{\eps_{z'}}\frac{\partial}{\partial z'}g_{H}^| &=\frac{1}{2}\begin{bmatrix}
\bar \kappa_{2}(e^{-\kappa_{2} \vert z-z' \vert}+\bar r_2 e^{-\kappa_{2}(z+z')}) &\bar \kappa_{3} (1-\bar r_{2})e^{-\kappa_{2} z+\kappa_{3} z'}\\
 \bar \kappa_{3}(1-\bar r_{2})e^{-\kappa_{2} z'+\kappa_{3} z} &\bar \kappa_{3} (e^{-\kappa_{3} \vert z-z' \vert}-\bar r_2 e^{\kappa_{3}(z+z')})\\
\end{bmatrix}\\
\tilde G_{yy}^{|}(k,z,z';\zeta)&=&\zeta^2 g_{E}^|&=\zeta^2\begin{bmatrix}
\frac{1}{2\kappa_2}(e^{-\kappa_{2}\vert z-z' \vert}-r_{2}e^{-\kappa_{2}(z+z')}) & \frac{1}{\kappa_2+\kappa_3}e^{\kappa_{3} z'-\kappa_{2} z} \\
 \frac{1}{\kappa_2+\kappa_3}e^{\kappa_{3} z-\kappa_{2} z'} & \frac{1}{2\kappa_3}(e^{-\kappa_{3}\vert z-z' \vert}+r_{2}e^{\kappa_{3}(z+z')})\\
\end{bmatrix}\nonumber\\
\tilde G_{zz}^{|}(k,z,z';\zeta)&=&-\frac{k^2}{\eps _z\eps_{z'}}g_{H}^|&=-k^2\begin{bmatrix}
\frac{1}{2 \eps_2\kappa_2}(e^{-\kappa_{2}\vert z-z' \vert}-\bar r_{2}e^{-\kappa_{2}(z+z')} )& \frac{1}{\eps_3\kappa_2+\eps_2\kappa_3}e^{\kappa_{3} z'-\kappa_{2} z} \\
\frac{1}{\eps_3\kappa_2+\eps_2\kappa_3} e^{\kappa_{3} z-\kappa_{z2} z'} &\frac{1}{2\eps_3\kappa_3}( e^{-\kappa_{3}\vert z-z' \vert}+\bar r_{2}e^{\kappa_{3}(z+z')})\\
\end{bmatrix}\nonumber\\
\tilde G_{xz}^{|}(k,z,z';\zeta)&=&-\frac{ik}{\eps_z\eps_{z'}}\frac{\partial}{\partial z}g_{H}^|&=\frac{ik}{2}\begin{bmatrix}
\frac{1}{\eps_{2}}(\text{sgn}(z-z')e^{-\kappa_{2} \vert z-z' \vert}-\bar r_2 e^{-\kappa_{2}(z+z')}) &\frac{1}{\eps_{3}}(1-\bar r_{2})e^{-\kappa_{2} z+\kappa_{3} z'}\\
-\frac{1}{\eps_{2}}(1+\bar r_{2})e^{-\kappa_{2} z'+\kappa_{3} z} &\frac{1}{\eps_{3}}(\text{sgn}(z-z')e^{-\kappa_{3} \vert z-z' \vert}-\bar r_2 e^{\kappa_{3}(z+z')})\\
\end{bmatrix}\nonumber\\
\tilde G_{zx}^{|}(k,z,z';\zeta)&=&\frac{ik}{\eps_z\eps_{z'}}\frac{\partial}{\partial z'}g_{H}^|&=\frac{ik}{2}\begin{bmatrix}
\frac{1}{\eps_{2}}(\text{sgn}(z-z')e^{-\kappa_{2} \vert z-z' \vert}+\bar r_2 e^{-\kappa_{2}(z+z')}) &\frac{1}{\eps_{2}}(1+\bar r_{2})e^{-\kappa_{2} z+\kappa_{3} z'}\\
-\frac{1}{\eps_{3}}(1-\bar r_{2})e^{-\kappa_{2} z'+\kappa_{3} z} &\frac{1}{\eps_{3}}(\text{sgn}(z-z')e^{-\kappa_{3} \vert z-z' \vert}+\bar r_2 e^{\kappa_{3}(z+z')})\\
\end{bmatrix}\nonumber\ .
\end{align}
The corresponding separation-dependent part is,
\begin{align}\label{Ga}
G_{xx}^{|a|}(k,z,z';\zeta, a)&=\frac{-\bar r_{1}}{2(e^{2a\kappa_{3}}-\bar r_{1}\bar r_{2})}\begin{bmatrix}
\bar \kappa_{2}(1-\bar r_{2}^2)e^{-\kappa_{2}(z+z')} &\bar\kappa_{3}( e^{-\kappa_{2} z}-\bar r_{2}e^{-\kappa_{2}z})( e^{-\kappa_{3} z'}-\bar r_{2}e^{\kappa_{3}z'})\\
 \bar\kappa_{3}( e^{-\kappa_{2} z'}-\bar r_{2}e^{-\kappa_{2}z'})( e^{-\kappa_{3} z}-\bar r_{2}e^{\kappa_{3}z}) &\bar\kappa_{3}( e^{-\kappa_{3} z}-\bar r_{2}e^{\kappa_{3}z})( e^{-\kappa_{3} z'}-\bar r_{2}e^{\kappa_{3}z'})\\
\end{bmatrix}\nonumber\\
G_{yy}^{|a|}(k,z,z';\zeta, a)&=\frac{\zeta^2r_{1}}{e^{2a\kappa_{3}}-r_{1}r_{2}}\begin{bmatrix}
\frac{1}{2\kappa_2}(1-r_{2}^2)e^{-\kappa_{2}(z+z')} & \frac{1}{\kappa_2+\kappa_3}(e^{-\kappa_{2} z-\kappa_{3} z' }+r_{2}e^{-\kappa_{2}z+\kappa_{3} z'})\\
  \frac{1}{\kappa_2+\kappa_3}( e^{-\kappa_{2} z'-\kappa_{3} z }+r_{2}e^{-\kappa_{2}z'+\kappa_{3} z}) & \frac{1}{2\kappa_3}(e^{-\kappa_{3} z}+r_{2}e^{\kappa_{3} z}) (e^{-\kappa_{3} z'}+r_{2}e^{\kappa_{3} z'})\\
\end{bmatrix}\\
G_{zz}^{|a|}(k,z,z';\zeta, a)&=\frac{-k^2\bar r_{1}}{e^{2a\kappa_{3}}-\bar r_{1}\bar r_{2}}\begin{bmatrix}
\frac{1}{2\eps_2\kappa_2}(1-\bar r_{2}^2)e^{-\kappa_{2}(z+z')} & \frac{1}{\eps_3\kappa_2+\eps_2\kappa_3}(e^{-\kappa_{2} z-\kappa_{3} z' }+\bar r_{2}e^{-\kappa_{2}z+\kappa_{3} z'})\\
 \frac{1}{\eps_3\kappa_2+\eps_2\kappa_3}( e^{-\kappa_{2} z'-\kappa_{3} z }+\bar r_{2}e^{-\kappa_{2}z'+\kappa_{3} z})& \frac{1}{2\eps_3\kappa_3}(e^{-\kappa_{3} z}+\bar r_{2}e^{\kappa_{3} z}) (e^{-\kappa_{3} z'}+\bar r_{2}e^{\kappa_{3} z'})\\
\end{bmatrix}\nonumber\\
G_{xz}^{|a|}(k,z,z';\zeta, a)&=\frac{i k\bar r_{1}}{2(e^{2a\kappa_{3}}-\bar r_{1}\bar r_{2})}\begin{bmatrix}
\frac{1}{\eps_{2}}(1-\bar r_{2}^2)e^{-\kappa_{2}(z+z')} &\frac{1}{\eps_{3}}( e^{-\kappa_{2} z}-\bar r_{2}e^{-\kappa_{2}z})( e^{-\kappa_{3} z'}+\bar r_{2}e^{\kappa_{3}z'})\\
 \frac{1}{\eps_{2}}( e^{-\kappa_{3} z}-\bar r_{2}e^{\kappa_{3}z})( e^{-\kappa_{2} z'}+\bar r_{2}e^{-\kappa_{2}z'}) &\frac{1}{\eps_{3}}( e^{-\kappa_{3} z}-\bar r_{2}e^{\kappa_{3}z})( e^{-\kappa_{3} z'}+\bar r_{2}e^{\kappa_{3}z'})\\
\end{bmatrix}\nonumber\\
G_{zx}^{|a|}(k,z,z';\zeta, a)&=\frac{-i k\bar r_{1}}{2(e^{2a\kappa_{3}}-\bar r_{1}\bar r_{2})}\begin{bmatrix}
\frac{1}{\eps_{2}}(1-\bar r_{2}^2)e^{-\kappa_{2}(z+z')} &\frac{1}{\eps_{2}}( e^{-\kappa_{2} z}+\bar r_{2}e^{-\kappa_{2}z})( e^{-\kappa_{3} z'}-\bar r_{2}e^{\kappa_{3}z'})\\
 \frac{1}{\eps_{3}}( e^{-\kappa_{2} z'}-\bar r_{2}e^{-\kappa_{2}z'})( e^{-\kappa_{3} z}+\bar r_{2}e^{\kappa_{3}z}) &\frac{1}{\eps_{3}}( e^{-\kappa_{3} z}+\bar r_{2}e^{\kappa_{3}z})( e^{-\kappa_{3} z'}-\bar r_{2}e^{\kappa_{3}z'})\\
\end{bmatrix}\nonumber
\end{align}

The limits of these propagators as $z$ and $z'$ approach $0$ are of particular interest. In this case the components of the matrices $ \tilde\bG^|(k;\zeta):= \tilde\bG^{|}(k,0,0;\zeta)$ and $\bG^{|a|}(k;\zeta,a):= \bG^{|a|}(k,0,0;\zeta,a)$ simplify to,
\begin{align}\label{Gsnew}
\tilde G_{xx}^{|}(k;\zeta)&=\frac{\kappa_2\kappa_3}{\eps_2\kappa_3+\eps_3\kappa_2}\begin{bmatrix}
1&1\\
 1&1\\
\end{bmatrix}&G_{xx}^{|a|}(k;\zeta,a)&=\frac{-\bar r_{1}(1-\bar r^2_2)\kappa_2}{2(e^{2a\kappa_{3}}-\bar r_{1}\bar r_{2})\eps_2}\begin{bmatrix}
1&1\\
1&1\\
\end{bmatrix}\\
\tilde G_{yy}^{|}(k;\zeta)&=\frac{\zeta^2}{\kappa_2+\kappa_3}\begin{bmatrix}
1&1 \\1&1\\
\end{bmatrix}&G_{yy}^{|a|}(k;\zeta,a)&=\frac{ r_{1}(1-r^2_{2})\zeta^2}{2(e^{2a\kappa_{3}}-r_{1}r_{2})\kappa_2}\begin{bmatrix}
1 &1\\
1& 1\\
\end{bmatrix}\nonumber\\
\tilde G_{zz}^{|}(k;\zeta)&=\frac{-k^2}{\eps_2\kappa_3+\eps_3\kappa_2}\begin{bmatrix}
\eps_3/\eps_2& 1\\
1 &\eps_2/\eps_3\\
\end{bmatrix}&G_{zz}^{|a|}(k;\zeta,a)&=\frac{-\bar r_{1}(1-\bar r^2_2)k^2}{2 (e^{2a\kappa_{3}}-\bar r_{1}\bar r_{2})\kappa_2\eps_3}\begin{bmatrix}
\eps_3/\eps_2 & 1\\
 1& \eps_2/\eps_3\\
\end{bmatrix}\nonumber\\
\tilde G_{xz}^{|}(k;\zeta)&=\frac{ik}{\eps_2\kappa_3+\eps_3\kappa_2}\begin{bmatrix}
\eps_{3}\bar\kappa_2 &\kappa_{2}\\
-\kappa_{3} &-\eps_2\bar\kappa_{3}\\
\end{bmatrix}&G_{xz}^{|a|}(k;\zeta,a)&=\frac{i \bar r_{1}(1-\bar r_{2}^2)k}{2(e^{2a\kappa_{3}}-\bar r_{1}\bar r_{2})}\begin{bmatrix}
1/\eps_{2} &1/\eps_{3}\\
 1/\eps_{2} &1/\eps_{3}\\
\end{bmatrix}\nonumber\\
\tilde G_{zx}^{|}(k;\zeta)&=\frac{-ik}{\eps_2\kappa_3+\eps_3\kappa_2}\begin{bmatrix}
\eps_3\bar\kappa_2 &-\kappa_3\\
\kappa_2 &-\eps_2\bar \kappa_3\\
\end{bmatrix}&G_{zx}^{|a|}(k;\zeta,a)&=\frac{-i \bar r_{1}(1-\bar r_2^2)k}{2(e^{2a\kappa_{3}}-\bar r_{1}\bar r_{2})}\begin{bmatrix}
1/\eps_{2} &1/\eps_{2}\\
 1/\eps_{3}&1/\eps_{3}\\
\end{bmatrix}\nonumber\ .
\end{align}

\section{Signed Correlators of the Roughness Profile}
\label{appB}

We here obtain the correlation functions of positive and negative components of the roughness profile for a Gaussian generating functional of roughness correlation functions,
\bel{GaussZ}
\vev{e^{\int d\bx \alpha(\bx)h(\bx)}}=e^{\half\int d\bx d\by \alpha(\bx) D_2(\bx-\by)\alpha(\by)}\ ,
\ee
that is fully determined by the two-point correlation function $\vev{h(\bx)h(\by)}=D_2(\bx-\by)$. We in the following assume that $D_2(0)\ge D_2(\bx-\by)>0$. 

Exploiting an integral representation of the $x\theta(x)$ distribution, one has that 
\bel{Heaviside}
h_\pm(\bx)=h(\bx)\theta(\pm h(\bx)) =\pm\frac{1}{2\pi}\lim_{\eps\rightarrow 0^+}\int_{-\infty}^\infty\frac{d \beta}{(\beta-i\eps)^2} e^{\pm i\beta h(\bx)}=\pm\lim_{\eps\rightarrow 0^+}\int_0^\infty \lambda d\lambda e^{-\eps\lambda} \int_{-\infty}^\infty\frac{d \beta}{2\pi} e^{-i\lambda\beta}e^{\pm i\beta h(\bx)}\ .
\ee
We use \equ{Heaviside} to write,
\bel{intrep}
\vev{h_+(\bx) h_\pm(\by)}=\pm\lim_{\eps\rightarrow 0^+}\int_0^\infty \lambda_1 d\lambda_1\int_0^\infty \lambda_2 d\lambda_2 e^{-\eps (\lambda_1+\lambda_2)} \int \frac{d\mathbf{\beta}}{(2\pi)^2} e^{-i\mathbf{\lambda}\cdot\mathbf{\beta}}\vev{e^{ i(\beta_1 h(\bx)\pm \beta_2 h(\by))}}
\ee
The expectation in\equ{intrep} is of the form given in \equ{GaussZ} with $\alpha(\bx')=i(\beta_1\delta(\bx'-\bx)\pm \beta_2 \delta(\bx'-\by))$ and therefore evaluates to,
\bel{evalvev}
\vev{e^{ i(\beta_1 h(\bx)\pm \beta_2 h(\by))}} = e^{-\half \mathbf{\beta}^T\cdot M_\pm \cdot\beta}\ ,
\ee
where the symmetric real, and positive $2\times 2$ matrix,
\bel{defM}
M_{\pm}=\begin{bmatrix} D_2(0) & \pm D_2(\bx-\by)\\ \pm D_2(\bx-\by) & D_2(0) \end{bmatrix}
\ee
has determinant $\det M_{\pm}=D^2_2(0)-D_2^2(\bx-\by)>0$ for $|\bx-\by|>0$. Performing the two-dimensional Gaussian integral in $\mathbf{\beta}=(\beta_1,\beta_2)$ (for $|\bx-\by|>0$) gives,
\bel{intbeta}
\vev{h_+(\bx) h_\pm(\by)}=\pm \frac{(\det M_\pm)^{-1/2}}{2\pi}\int_0^\infty  d\lambda_1\int_0^\infty  d\lambda_2 \lambda_1\lambda_2 e^{-\half \mathbf{\lambda}^T\cdot M^{-1}_\pm \cdot\lambda} \ .
\ee
Converting to polar coordinates $(\lambda_1,\lambda_2)=\lambda (\cos\theta,\sin\theta)$ and noting that the integral extends over the first quadrant with $0<\theta<\pi/2$ only,  
\begin{align}\label{intbeta1}
\vev{h_+(\bx) h_\pm(\by)}&=\pm \frac{(\det M_\pm)^{-1/2}}{2\pi}\int_0^{\pi/2} d\theta  \frac{\sin(2\theta)}{2}  \int_0^\infty \lambda^3 d\lambda  e^{-\half \lambda^2 (D_2(0)\mp \sin(2\theta) D_2(\bx-\by))/\det M_{\pm}}\nonumber\\
&=\pm \frac{(\det M_\pm)^{3/2}}{4\pi}\int_0^{\pi} d\theta  \frac{\sin\theta}{(D_2(0)\mp  D_2(\bx-\by)\sin\theta)^2}\nonumber\\
& =\pm\frac{D_2(0)}{2\pi}(\sin\phi+(\frac{\pi}{2}\pm \frac{\pi}{2} -\phi)\cos\phi)
\end{align}
with $\cos\phi=D_2(\bx-\by)/D_2(0), 0<\phi<\pi/2$. This result is reproduced in \equ{Dsigned}. The last expression uses that the lengths $D_2(0), D_2(\bx-\by) $ and $\det M_\pm$ can be interpreted as the sides of a right triangle with hypotenuse $D_2(0)$. 

\section{Angular Integrals}
\label{appC}
For the class of  correlations functions,
\bel{affine}
D_s(q)=2\pi\sigma^2 l_c^2(1+q^2 l_c^2/(2 s))^{-1-s} \ \ \text{with}\ \ s>0,
\ee
the angular integrals of Eqs.~(\ref{fig2b1}),~(\ref{fig2b1}),~(\ref{fig2b1})~and~(\ref{fig2b1}) are all of the form,
\bel{angular}
A_n(s)=\int_{-\pi}^\pi  \frac{d\theta\cos^n\theta}{(1+a-b\cos\theta)^{s+1}} =\frac{\Gamma(s+1-n)}{\Gamma(s+1)}\frac{\partial^n}{\partial b^n}\frac{2\pi }{ (1+a+b)^{s+1-n}}\ {_2F_ 1}(\half,s+1-n;1;\frac{2 b}{1+a+b})\ ,
\ee
with $a=\half (k^2+k'^2) l_c^2/s \geq b=k k' l_c^2/s>0$ and $n=0,1,2$. They are given by values of the generalized hypergeometric function ${_2F_ 1}(\half,\nu;1;x)$ for any $s>0$. 

The exponential roughness correlation $D_\text{Exp}$ of \equ{ExpD} corresponds to $s=1/2$ and the relevant angular integrals in this case are complete elliptic integrals, 
\begin{align}\label{angExp}
A_0(1/2)&=\frac{4}{(1+a-b)\sqrt{1+a+b}} E(\frac{2 b}{1+a+b})\nonumber\\
A_1(1/2)&=\frac{4}{(1+a-b)b\sqrt{1+a+b}} \left((1+a) E(\frac{2 b}{1+a+b})-(1+a-b)K(\frac{2 b}{1+a+b})\right)\nonumber\\
A_2(1/2)&=\frac{4}{(1+a-b)b^2\sqrt{1+a+b}} \left((2(1+a)^2-b^2) E(\frac{2 b}{1+a+b})-2(1+a)(1+a-b)K(\frac{2 b}{1+a+b})\right),
\end{align}
with $a=(k^2+k'^2) l_c^2$ and $b=2 k k' l_c^2$.

The limit $s\rightarrow\infty$ of the Gaussian correlation in \equ{GaussD} is best obtained directly. The angular integrals in this limit are,
\bel{angularG}
A_n(\infty)=e^{-l_c^2(k^2+k'^2)/2}\int_{-\pi}^\pi  d\theta\cos^n\!\theta \ e^{l_c^2 k k'\cos\theta} =2 \pi e^{-l_c^2(k^2+k'^2)/2} \left.\frac{\partial^n}{\partial \alpha^n} I_0(\alpha)\right|_{\alpha=l_c^2 k k'},
\ee
where $I_0(x)$ is the modified Bessel function of the first kind of $0$-th order. The relevant angular integrals for Gaussian roughness correlation thus are,
\begin{align}\label{angGauss}
A_0(\infty)&=2 \pi e^{-l_c^2(k^2+k'^2)/2}I_0(l_c^2 k k') \nonumber\\
A_1(\infty)&=2 \pi e^{-l_c^2(k^2+k'^2)/2}I_1(l_c^2 k k') \nonumber\\
A_2(\infty)&=\pi e^{-l_c^2(k^2+k'^2)/2}(I_0(l_c^2 k k')+I_2(l_c^2 k k'))\  
\end{align}

\section{The Response Function}
\label{appD}
The roughness correction to the Casimir free energy of order $\sigma^2$ is given in\equ{rough2F}. This correction is linear in $D(q)$ and one may  define\cite{Neto2005} the response function $R_T(q, a)$ of \equ{defR} defined by,
\begin{align}\label{correction}
\Delta F_T^{Cas}[a]&=\half\vev{\tr  \tilde\bV^h\bG^{|a|}}-\half\vev{\tr \tilde\bV^h\tilde \bG^|\tilde\bV^h \bG^{|a|}}+\half\tr  \delta\tilde\bV^h\bG^{|a|} -{\textstyle \frac{1}{4}}\vev{\tilde \bV^h\bG^{|a|}\tilde\bV^h\bG^{|a|}}\nonumber\\
&=\int_0^\infty \frac{q dq}{2\pi}D(q) R_T(q, a) \
\end{align}
To obtain $R_T(q,a)$ we change the integration variable from $\bk'$ to $\bq=\bk'-\bk$ in Eqs.~(\ref{seagull}),~(\ref{fig2b1}),~(\ref{fig2c})~and~(\ref{fig2d}) and choose $\bk=(k,0)$ to define the positive $x$-axis. In these coordinates $k'_x=k+q\cos\theta, k'_y=q\sin\theta$ and explicit expressions for the response function $R_T(q,a)$ can be read off from,
\begin{subequations}
\begin{align}  
\half\vev{\tr  \tilde\bV^h\bG^{|a|}}&=\int_0^\infty \frac{q dq}{2\pi}D(q)\sum_n(-AT)\int_0^\infty\frac{k d k}{2\pi} \kappa\kappa_\eps ( \frac{\bar r^2}{e^{2 a \kappa}-\bar r^2} +\frac{ r^2}{e^{2 a \kappa}- r^2})\ , \label{seagulT0}\\
-\half\vev{\tr \tilde\bV^h\tilde \bG^|\tilde\bV^h \bG^{|a|}}&=\int_0^\infty \frac{q dq}{(2\pi)^2}D(q)\sum_n(-AT)(\eps-1)^2\int_0^\infty\frac{k d k}{2\pi}\int_{-\pi}^{\pi} d\theta\Big[\frac{r (1-r^2) \zeta^2 }{4(e^{2 a \kappa}-r^2)\kappa_\eps}\times\label{fig2bT0}\\
&\hspace{-9em}\left(\frac{\kappa'\kappa'_\eps}{\eps\kappa'+\kappa'_\eps}\Big(\frac{k'_y}{k'}\Big)^2+\frac{\zeta^2}{\kappa'+\kappa'_\eps}\Big(\frac{k'_x}{k'}\Big)^2\right)+\frac{ \bar r(1-\bar r^2)}{4 (e^{2 a\kappa}-\bar r^2)\eps}\left(\frac{\eps k^2 k'^2}{\kappa_\eps(\eps\kappa'+\kappa'_\eps)}-kk'_x\bar r'- \frac{\kappa_\eps\kappa'\kappa'_\eps}{\eps\kappa'+\kappa'_\eps}\Big(\frac{k'_x}{k'}\Big)^2- \frac{\kappa_\eps\zeta^2}{\kappa'+\kappa'_\eps}\Big(\frac{k'_y}{k'}\Big)^2 \right)\Big]\ \nonumber \\
-{\textstyle \frac{1}{4}}\vev{\tr \tilde\bV^h\tilde \bG^{|a|}\tilde\bV^h \bG^{|a|}}&=\int_0^\infty \frac{q dq}{(2\pi)^2}D(q)\sum_n(-AT)(\eps-1)^2\int_0^\infty\frac{k d k}{2\pi}\int_{-\pi}^{\pi} d\theta\Big[\frac{r (1-r^2) \zeta^2 }{16(e^{2 a \kappa}-r^2)\kappa_\eps}\label{fig2b2T0}\\
&\hspace{-8em}\left(\frac{r' (1-r'^2) \zeta^2}{(e^{2 a \kappa'}-r'^2)\kappa'_\eps}\Big(\frac{k'_x}{k'}\Big)^2-\frac{2\bar r'(1-\bar r'^2)\kappa'_\eps}{(e^{2a\kappa'}-\bar r'^2)\eps}\Big(\frac{k'_y}{k'}\Big)^2\right)+\frac{ \bar r\bar r'(1-\bar r^2)(1-\bar r'^2)}{16 (e^{2 a\kappa}-\bar r^2)(e^{2 a\kappa'}-\bar r'^2)}\left(\frac{k^2 k'^2}{ \kappa_\eps\kappa'_\eps}+\frac{2 k k'_x}{\eps }+\frac{\kappa_\eps\kappa'_\eps}{\eps^2 }\Big(\frac{k'_x}{k'}\Big)^2 \right)\Big]\  ,\nonumber\\
\half\tr \delta\tilde\bV \bG^{|a|}&=\int_0^\infty \frac{q dq}{2\pi}D(q)\sum_n AT(\eps-1)^2\int_0^\infty\frac{k d k}{2\pi}\Big[\frac{ \bar r(1-\bar r^2)}{ 4(e^{2 a\kappa}-\bar r^2)}\frac{k^2q^2}{\kappa_{\eps}(\eps\kappa'+\kappa'_\eps)}+\nonumber\\
&\hspace{2em}\left(\frac{r (1-r^2) \zeta^2 }{4(e^{2 a \kappa}-r^2)\kappa_\eps}-\frac{ \bar r(1-\bar r^2)\kappa_\eps}{4(e^{2 a\kappa}-\bar r^2)\eps}\right)\left(\frac{\kappa'\kappa'_\eps/2}{\eps\kappa'+\kappa'_\eps}+\frac{\zeta^2/2}{\kappa'+\kappa'_\eps}-\frac{g^2 \zeta}{1 +\sqrt{\eps}}\right)\Big]\ .\label{fig2cT0}
\end{align}
\end{subequations}
In the last (counter term) expression of \equ{fig2cT0} $\kappa'=\sqrt{\bq^2+\zeta^2}$ and $\kappa'_\eps=\sqrt{\bq^2+\zeta^2\eps(\zeta)}$ . Note that the angular integration in these coordinates cannot be performed analytically.


\bibliography{%
       biblio/b2014-roughness%
              }

\end{document}